\documentclass[10pt]{iopart}

\usepackage{iopams} 
\usepackage{mathrsfs} 

\usepackage{graphicx}
\usepackage[dvipsnames]{xcolor} 
\usepackage{caption}

\usepackage{booktabs}
\usepackage[numbers,sort&compress]{natbib}
\usepackage{url,hyperref}
\hypersetup{colorlinks=true,
linkcolor=blue,
citecolor=purple,
filecolor=purple,
urlcolor=blue}

\usepackage{setspace} 
\AtBeginDocument{\setstretch{1.125}}

\usepackage{outlines} 
\usepackage[splitrule]{footmisc} 

\usepackage{tensor} 
\usepackage{array}

\usepackage{slashed}
\usepackage{cancel}
\usepackage{verbatim}
\usepackage{csquotes}

\usepackage{amstext} 
\newcolumntype{C}{>{$}c<{$}} 
\newcolumntype{L}{>{$}l<{$}} 
\usepackage{multirow}

\newtheorem{theorem}{Theorem}[section]

\newtheorem{definition}[theorem]{Definition}
\newtheorem{conjecture}[theorem]{Conjecture}

\newtheorem{proposition}[theorem]{Proposition}

\renewcommand{\theequation}{\arabic{section}.\arabic{subsection}.\arabic{equation}}
\newcommand{\cleqn}{\setcounter{equation}{0}}

\begin{document}

\title[On the status of Strong Cosmic Censorship]{A note on Strong Cosmic Censorship and its violation in Reissner-Nordstr\"om
de Sitter black hole space-times
}

\author{Anna Chrysostomou$^{1}$, Alan S. Cornell$^{2}$, Aldo Deandrea$^{3,2}$, and Seong Chan Park$^4$}
\address{$^{1}$Laboratoire de Physique Th\'eorique et Hautes \'Energies - LPTHE, Sorbonne Universit\'e, CNRS, 4 Place Jussieu, 75005 Paris, France
}
\address{$^{2}$Department of Physics, University of Johannesburg, PO Box 524, Auckland Park 2006, South Africa}
\address{$^{3}$Universit{\'e} Claude Bernard Lyon 1, IP2I, UMR 5822, CNRS/IN2P3, 4 rue Enrico Fermi, 69622 Villeurbanne Cedex, France}
\address{$^4$Department of Physics and IPAP, Yonsei University, Seoul 03722, Republic of Korea}

\eads{\mailto{chrysostomou@lpthe.jussieu.fr}, acornell@uj.ac.za, deandrea@ip2i.in2p3.fr, and sc.park@yonsei.ac.kr}

\vspace{10pt}

\begin{abstract}
Penrose's Strong Cosmic Censorship conjecture safeguards determinism in General Relativity. Within the initial value approach to General Relativity, proof of strong cosmic censorship preservation is predicated on the unique evolution of the metric. For the Kerr-Newman family of black hole solutions, this requires the inextendability of the metric past the Cauchy horizon, due to the development of a \enquote{blue-shift} instability. Attempts to provide a rigorous mathematical proof of strong cosmic censorship has led to the formulation of several strong cosmic censorship conjectures of varying strengths, which seem to be discussed rarely outside of the mathematical relativity literature. In this note, we review some of the arguments for and against strong cosmic censorship preservation, with a focus on the Reissner-Nordstr\"om de Sitter context, where the positive cosmological constant invites a \enquote{red-shift} effect that competes against the \enquote{blue-shift}. We study the consequent role of quasinormal mode behaviour and illustrate the parameter space for which we consistently observe violations of the strong cosmic censorship conjecture within Reissner-Nordstr\"om de Sitter black holes. 
\end{abstract}

\section{Introduction \label{sec:SCC}}\cleqn

\par Within the framework of General Relativity (GR), a curvature singularity exists at $r=0$. For much of the twentieth century, this was accepted as unphysical: Einstein and his contemporaries considered the singularity to be an artefact of the theory, while Oppenheimer and Snyder \cite{OppenheimerSnyder1939_StarCollapse} suggested it to be a consequence of imposing unrealistic symmetry idealisations on stellar collapse (see Ref. \cite{Landsman2022_PenroseContextualised} for a comprehensive historical review). The 1960s, however, saw a shift in the paradigm, beginning with Penrose's seminal 1965 paper on gravitational collapse \cite{Penrose1964_CC1}. There, he demonstrated that in the wake of the inward stellar collapse to $r=0$ and the subsequent formation of a trapped surface (\textit{viz.} a surface from which light cannot escape outwards), a space-time $(\mathcal{M},g)$ cannot be future null-geodesically complete. In other words, a light-like trajectory incoming from spatial infinity terminates at an undefined point at $r=0$. With this, a vague suggestion of a physical manifestation of the singularity came to be. 

\par In isolation, Ref. \cite{Penrose1964_CC1} makes no claims on the nature of the singularity nor on the implications of its existence. To do so would have been impossible, since its publication preceded the definition of an event horizon and, by extension, a black hole. In Ref. \cite{Penrose1969_CC2-WCCdef}, however, Penrose provides the outstanding definition of the black hole event horizon as the absolute boundary of the set of all events (i.e. the space-like boundary of the past of future null infinity $\mathscr{I}^+$), corresponding to the coordinate singularity $r=r_+$, which can be observed in principle by an external inertial observer. The black hole is then defined as the region bounded by the event horizon (see  \ref{app:conform} for details). 
\par With these definitions in place, the formulation of the cosmic censorship conjecture could follow, forbidding naked singularities within GR \cite{Penrose1969_CC2-WCCdef}. Today, we refer to this requirement for cloaked singularities as the \enquote{Weak Cosmic Censorship} (WCC) conjecture. Despite the absence of a formalised proof, WCC serves as an implicit assumption throughout the GR literature, having become the cornerstone of black hole uniqueness theorems and models. As we shall demonstrate in Section \ref{sec:RNdSBH}, the WCC conjecture is crucial to our understanding of the black hole structure and phase space. 

\par In this note, however, we are concerned with the notion of \enquote{Strong Cosmic Censorship} (SCC),\footnote{While this follows the GR convention of referring to conjectures as \enquote{strong} or \enquote{weak}, this does not indicate the relative strength of one to another. In fact, the link between the two is demonstrably tenuous (see Ref. \cite{Landsman2022_PenroseContextualised} for further discussion).} with a particular interest in providing a pedagogical overview of how the problem is defined and approached in the literature. The underlying principle of the SCC conjecture lies in the inextendability of a physically-reasonable space-time metric past the space-like $r=0$ singularity \cite{Penrose1964_CC1,Penrose1999_CC}. Intuitively, we may consider an infalling object whose journey past the event horizon must end at $r=0$. To define this formally, however, is highly non-trivial, as we need to specify the criteria that must be satisfied for a space-time to be considered \enquote{physically-reasonable} as well as the requirements for \enquote{inextendability}. 

\par Further complications arise when we consider the Kerr-Newman family of black holes: the region bounded by the event horizon includes an additional Killing horizon, the Cauchy horizon, as well as a time-like singularity. The criteria for \enquote{physically-reasonable} must be adapted to accommodate this structure. Worse still, SCC preservation becomes conceptually entangled with determinism in this setup: an observer that crosses the Cauchy horizon can, upon looking back whence they came, see the entire future of the asymptotically-flat spacetime exterior to the black hole within a finite time.  

\par In other words, the Cauchy horizon marks the limit where initial data evolves uniquely. Within the GR framework, only if the metric is inextendible past this horizon can there be a possibility to rescue SCC. Upon closer inspection, the Cauchy horizon is in fact demonstrably unstable. Infalling radiation becomes infinitely \enquote{blue-shifted} (i.e. its oscillation frequency increases) as it approaches $r=r_-$ and accumulates at the Cauchy horizon. This leads to a curvature divergence. As such, this \enquote{blue-shift} mechanism results in a singularity, preventing causal curves from being extended past $r=r_-$ \cite{Penrose1973_Blueshift}. The criterion for whether this holds true in a space-time with a non-zero cosmological constant shall be the focal point of our discussion.

\par Cosmic censorship in general continues to draw interest, with recent investigations concerning its relationship to Swampland conjectures \cite{HorowitzCrisford2017_WCCtesting} and especially possible violations in the presence of a non-zero cosmological constant (see Ref. \cite{Rossetti2023_SCCmath} and references therein). The SCC conjecture itself has been challenged and adapted several times; the literature on the subject is scattered across multiple sub-disciplines, and includes philosophical deconstructions, rigorous mathematical arguments, and numerical analyses. According to the existing definitions for the singularity and the horizon, the preservation of SCC is predicated on the inextendability of the space-time under specific conditions. For this reason, the arguments in favour of and in opposition to SCC have become dominated by the \enquote{initial-value approach} to GR, also referred to as the \enquote{partial differential equations programme}. In this approach, SCC preservation is posed as an initial-value problem that seeks to prove whether for \enquote{generic} initial data, the maximal globally hyperbolic data is inextendible as a solution to the vacuum Einstein equations \cite{Landsman2022_PenroseContextualised}. This allows for the establishment of specific criteria that, once met, formally proves that SCC holds. In this manner, six decades after its formulation, a formal proof of the $(C^0\text{-})$inextendibility of the Schwarzschild space-time was presented \cite{Sbierski:2015nta}.

\par With these perspectives in mind, we produce a short and pedagogical treatment of SCC. We outline the principles that motivated the initial formulation of the conjecture, addressing several of the most recent revisions to the statement, and clarifying the distinctions between the different formulations of the SCC conjecture. Our discussion culminates in the Reissner-Nordstr\"om de Sitter (RNdS) black hole space-time, which is the central focus of this note. Due to the presence of both Cauchy and cosmological horizons, the evolution of initial data across the event horizon and into the RNdS black hole interior is particularly interesting. To explore the influence of mass and charge on the evolution of a scalar field within the RNdS black hole interior, we consider a complex scalar field endowed with mass $\mu$ and charge $q$ as a proxy for the full nonlinear Einstein field equations. 

\par Our note is structured as follows: To contextualise our discussion, we begin with a brief review of the RNdS space-time in Section \ref{sec:RNdSBH}. Specifically, we describe the RNdS black hole structure and the phase space of RNdS black hole solutions in Section \ref{subsec:RNdSstructure}. In Section \ref{subsec:RNdSQNF}, we introduce the massive charged scalar test field oscillating within the RNdS black hole space-time and introduce the quasinormal modes (QNMs) - the damped perturbations in space-time that provide characteristic information about their black hole source.     From semi-classical considerations of a charged and massive scalar field propagating in the RNdS background, we demonstrate that there are regions in the RNdS parameter space for which we observe violations of the SCC conjecture. Our review of the SCC conjecture is contained in Section \ref{sec:SCCreview}, where we begin with the Schwarzschild black hole and build up to the RNdS. This review is targeted towards the non-expert in mathematical relativity, in order to familiarise the reader with the literature on the subject. It is for this reason that we provide the supplementary  \ref{app:GRdefs} and \ref{app:conform}. Finally, in Section \ref{subsec:SCCinRNdS}, we demonstrate through a semi-classical analysis of the QNM behaviour at the event horizon that we consistently find evidence for violations of SCC in the extremised region of the RNdS black hole space-time. The final section is dedicated to conclusions.

\par To facilitate our discussion on the formulation of the SCC conjecture within the RNdS black hole space-time, we must first identify the characteristics of the space-time. Furthermore, since the evolution of a charged and massive scalar field oscillating in the background of the RNdS black hole space-time will form a central component of our analysis, we introduce it as well as its QNM behaviour in the next section.

 \section{The Reissner-Nordstr{\"o}m de Sitter black hole \label{sec:RNdSBH}}\cleqn
 \subsection{RNdS black hole background \label{subsec:RNdSbackground}}\cleqn
\par To examine the $(3+1)$-dimensional RNdS black hole, let us consider the Einstein-Hilbert-Maxwell action in an asymptotically-de Sitter space-time, closely following the discussion we provided in Ref. \cite{Chrysostomou:2024inc}. Under Planck units,
\begin{equation} \label{eq:action}
S =  \int d^4x \sqrt{-g} \left[ \frac{1}{16 \pi G}\left(R - 2\Lambda  \right) - \frac{1}{4g_1^2} F_{\mu \nu} F^{\mu \nu} \right] \;,
\end{equation}  
\noindent where $\kappa^2~=~8 \pi G~=~M^{-2}_P$ relates the gravitational coupling $\kappa$ to Newton's gravitational constant $G$ and the Planck mass $M_P$. The de Sitter radius $L_{dS}$, Hubble parameter $H$, and cosmological constant $\Lambda >0$ can be expressed as $H^2~=~L^{-2}_{dS}~=~\Lambda /3$. Here, we are concerned only with the electrically-charged RNdS black hole, such that the only non-zero component of the electromagnetic field strength tensor $F^{\mu \nu}$ is
\begin{equation}
F_{tr} = \frac{g_1^2}{4 \pi} \frac{q_{_{_{\rm BH}}}}{r} 
\;,
\end{equation}
for a $U(1)$ coupling $g_1$. Per the \enquote{no-hair conjecture} \cite{Israel1968_NoHair-RN}, the black hole is completely specified by its mass $m_{_{_{\rm BH}}}$ and charge $q_{_{_{\rm BH}}}$. The geometry of the space-time is encoded in the Ricci scalar curvature $R~=~g^{\mu \nu} R_{\mu \nu}$ and the metric $g~=~\det |g_{\mu \nu}|$. 

\par The Lagrangian of Eq. (\ref{eq:action}) admits the static and spherically-symmetric black hole solution \cite{Kramer2003_ExactEFEsols}, 
\begin{equation} \label{eq:metric}
ds^2 = -f(r) dt^2 \; + \; f(r)^{-1} dr^2 \; + \; r^2 \left(d\theta^2 +  \sin^2 \theta  d \phi^2 \right) \;, 
\end{equation}
\noindent written in terms of the Schwarzschild coordinates $(t,r,\theta,\phi)$, with $t \in (-\infty,+\infty)$, $\theta \in (0,\pi),$ and $\phi \in (0,2\pi)$. The metric function $f(r)$ reduces to
\begin{equation} \label{eq:f}
    f(r) = 1 - \frac{2M}{r} + \frac{Q^2}{r^2} - \frac{r^2}{L^2_{dS}}  \;,
\end{equation}
\noindent under the parametrisations \cite{AntoniadisBenakli2020_WGCdS}:
\begin{eqnarray}
\hspace{1.8cm}Gm_{_{\rm BH}} = \frac{\kappa^2}{8 \pi}m_{_{\rm BH}} &=& M \;, \\
\hspace{0.2cm}\frac{G}{4 \pi r^2}g^2_1 q_{_{\rm BH}} = \frac{\kappa^2}{32 \pi^2 r^2}g^2_1 q_{_{\rm BH}}&=& Q^2 \;, \\
\hspace{4cm}\frac{3}{\Lambda} &=& L^2_{dS} \;.
\end{eqnarray}
\noindent From the roots of Eq. (\ref{eq:f}), we obtain the locii of the horizons of the RNdS black hole. The causally-connected regions of the space-time are delineated accordingly. Furthermore, since the WCC conjecture forbids the exposure of the curvature singularity, the construction of the parameter space of valid RNdS black hole solutions is predicated on the nature of these roots. We shall elaborate on these in the next subsection.

\subsection{RNdS black hole structure and (M,Q) phase space \label{subsec:RNdSstructure}}\cleqn
\par The global structure of the black hole space-time is predicated on the nature at the horizons and at the boundaries (i.e. the scalar curvature singularity $r \rightarrow 0$ and, in asymptotically flat space-time, spatial infinity $r \rightarrow \infty$). The roots of the metric function dictate the causal structure of the space-time, and depend strongly on the values of $M$, $Q$, and $L_{dS}$. From the four real roots of Eq. (\ref{eq:f}), we can identify three Killing horizons: the Cauchy horizon $r=r_-$, the event horizon $r=r_+$, and the cosmological horizon $r=r_c$, where
\begin{equation} \label{eq:RNdShorizons}
0 < r_- \leq r_+ \leq r_c \leq L_{dS} < \infty \;.
\end{equation}
\noindent The fourth (and unphysical) root is given by $r_0 = -(r_- + r_+ + r_c)$. These allow for an alternate expression of the metric function, 
\begin{equation} \label{eq:fhor}
f(r) = \frac{1}{r^2 L^2_{dS}}(r-r_-)(r-r_+)(r_c-r)(r-r_0) \;.
\end{equation}

\begin{figure}[t]
    \centering
    \includegraphics[width=0.8\linewidth]{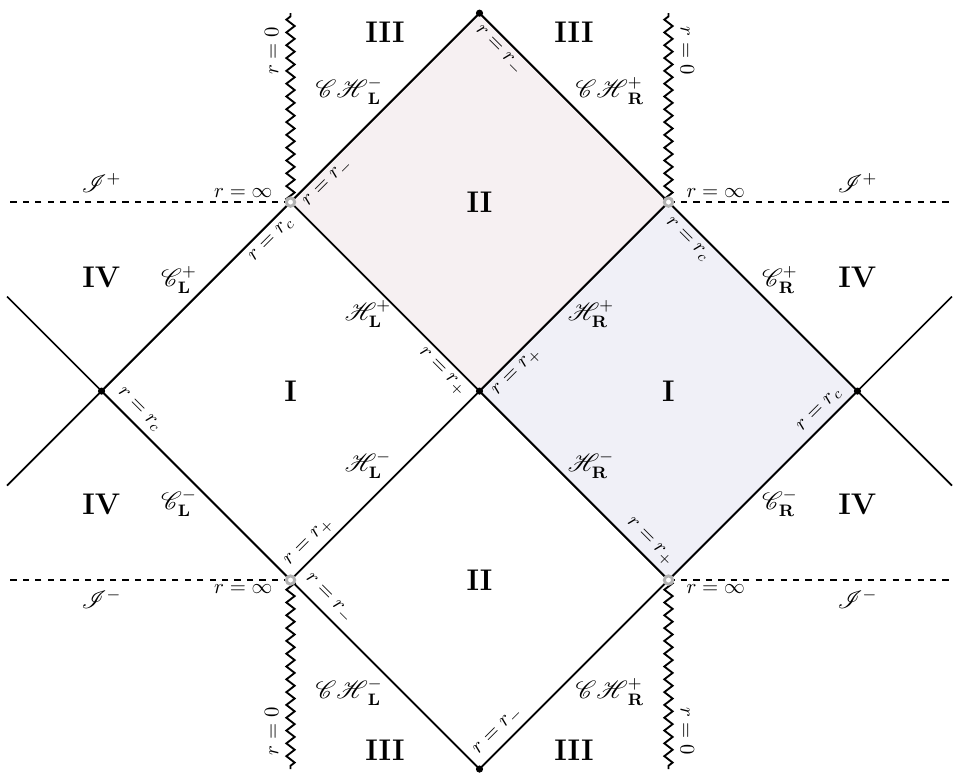}
    \vskip 0.3cm
    \caption{\textit{The conformal diagram for the extended RNdS space-time. Shading indicates the black hole interior (Region II) and the accessible universe (Region I). 
}}
    \label{fig:Penrose_RNdS}
\end{figure}

\par The physical horizons of Eq. (\ref{eq:RNdShorizons}) divide the black hole space-time into four causally-connected regions. In Fig. \ref{fig:Penrose_RNdS}, we sketch the extended conformal space-time diagram for the RNdS black hole, following the conventions set out in \ref{app:conform}. Respectively, these are the static patch $\mathcal{M}_{ext}$ (Region I), bounded by the bifurcate event horizon $\mathscr{H}^+_R \cup \mathscr{H}^-_R$ and the cosmological horizon $\mathscr{C}^+_R$, and $\mathcal{M}_{int}$ (Region II), bounded by the bifurcate event horizon $\mathscr{H}^+_L \cup \mathscr{H}^+_R$ and the Cauchy horizon $\mathscr{CH}^+_R$. We discriminate between \enquote{ingoing} (right) and \enquote{outgoing} (left) horizons using the subscripts $R$ and $L$, respectively; we are only focused on the ingoing horizons. We use shading to highlight the RNdS black hole interior (Region I) and the observable universe (Region II). In Table \ref{table:RNdSdom}, we summarise the dominant contribution from $f(r)$ in each region of the space-time \cite{AntoniadisBenakli2020_WGCdS}: 
\begin{table}[h]
\caption{\textit{RNdS space-time regions (Fig. \ref{fig:Penrose_RNdS}) and the dominant force within each.  \label{table:RNdSdom}}}
\begin{center}
\begin{tabular}{@{}CCCCC@{}}
\hline\noalign{\smallskip}
\hfill Region & & \hfill f(r) &  & Dominant \; force \\
\noalign{\smallskip}\hline\noalign{\smallskip}
0 < r < r_- & (III) & \sim Q^2/r^2 &>0 & \textit{repulsive electromagnetic force}  \\
r_- < r < r_+ & (II)& \sim -2M/r &< 0 & \textit{attractive gravitational force} \\
r_+ < r < r_{c} & (I) & \sim 1 &>0 & \textit{no subdominant contributions}\\
r_{c} < r & (IV) &  \sim - r^2 &< 0 & \textit{positive vacuum energy} \\
 \hline\noalign{\smallskip}
\end{tabular}
\end{center}
\end{table}

\par For each non-degenerate horizon specified in Eq. (\ref{eq:RNdShorizons}), we can calculate a surface gravity $\kappa_i$ (see \ref{app:conform} for details) and an associated Hawking temperature $T_i$ \cite{Hawking1974_HawkT}. For $i \in \{-,+,c\}$, we define these as
\begin{equation}
\kappa_i = \frac{1}{2}  \frac{d}{dr} f(r) \bigg \vert_{r=r_i} \;, \quad \quad T_i = \frac{\kappa_i}{2 \pi} \;.
\end{equation}
\noindent Explicitly, the surface gravities corresponding to each horizon are:
\begin{eqnarray}
\kappa_- & =  & - \frac{(r_+ - r_-)(r_c - r_-)(r_- - r_0)}{2r_-^2 L^2_{dS}} \label{eq:km} \;, \\
\kappa_+ & =& + \frac{(r_+ - r_-)(r_c - r_+)(r_+ - r_0)}{2r_+^2 L^2_{dS}} \label{eq:kp}  \;, \\
\kappa_c & = & -\frac{(r_c - r_-)(r_c - r_+)(r_c - r_0)}{2r_c^2 L^2_{dS}} \;, \label{eq:kc}  
\end{eqnarray}
\noindent with $|\kappa_-| > |\kappa_+|$ \cite{MossMyers1998_CC}. Generally, the Hawking radiation emanates from each horizon at a different temperature.

\par While thermal radiation isotropically pervades de Sitter space-time \cite{GibbonsHawking1977_BHTherm}, the black hole is not necessarily in thermal equilibrium with the de Sitter edge. In fact, there are just two families of black hole solutions that represent thermal equilibrium: the \enquote{lukewarm} and the \enquote{charged Nariai} families of RNdS black hole solutions. The former represents the case for which $M=Q$, while the latter corresponds to the condition $r_+ = r_c$ \cite{Romans1991_ColdLukewarmRNdS}. 

\par
\begin{figure}[t!]
\centering
\includegraphics[width=0.6\linewidth]{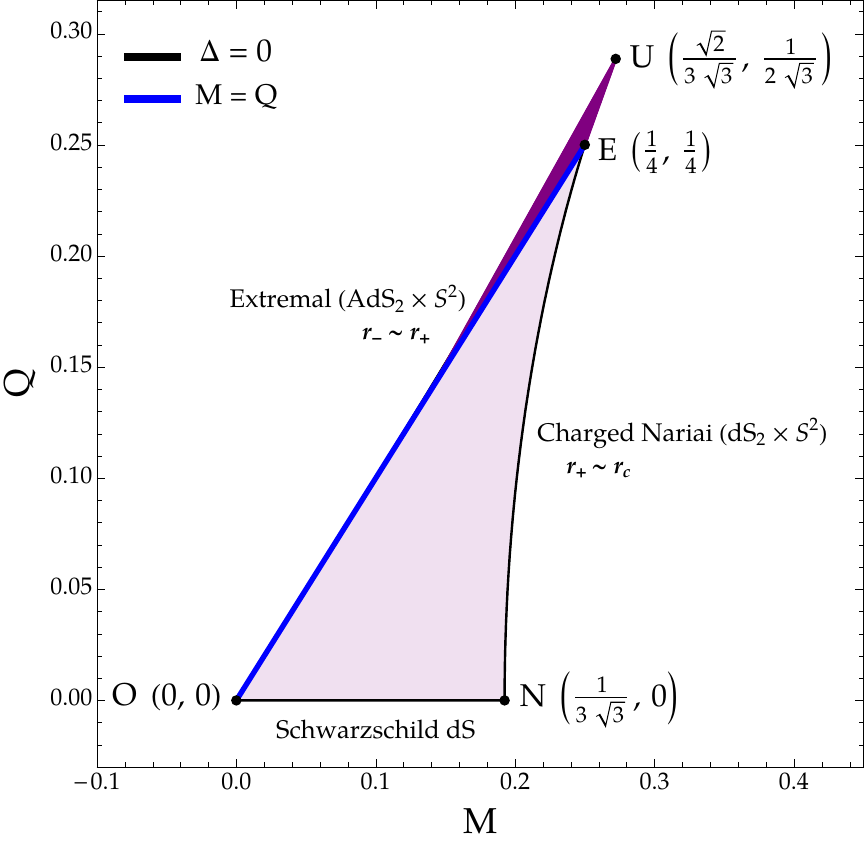}
\caption{ \textit{A two-dimensional projection of the parameter space for $H^2 = \Lambda/3 =1$ for the 4D RNdS black hole. Dark (light) shading corresponds to cold (warm) black holes. Diagram originally sketched in Ref. \cite{Chrysostomou:2024inc}, inspired by Ref. \cite{vanRiet2019_FLevapBHdS}.
}}
\label{fig:sharkfin}
\end{figure}

 These correspond to non-extremal black holes. However, degenerate horizons have vanishing surface gravities. Since $T_i=0$ in these cases, these black holes are classified as \enquote{cold}. For asymptotically flat space-time, the Reissner-Nordstr{\"o}m (RN) black hole develops a degenerate horizon when $M=Q$.\footnote{Recall that to avoid superextremality, $M/Q \leq 1$ for RN black holes.} As we can illustrate explicitly in Fig. \ref{fig:sharkfin}, the presence of the cosmological constant affects the $(M,Q)$ parameter space in the RNdS case: there exists a region in the parameter space of subextremal black hole solutions for which $Q>M$. For the RNdS black hole, $\kappa_- = \kappa_+ = 0$ when $Q_{ext} \sim r_+$, as suggested in Ref. \cite{DiasReallSantos2018_SCC} and shown in Ref. \cite{Chrysostomou2023_RNdS} particularly for low $Q$. The definition of $Q_{ext}$ is provided in Ref. \cite{DiasReallSantos2018_SCC}, $viz.$
\begin{equation} \label{eq:qext}
Q_{ext} \equiv y_+ r_c \sqrt{\frac{1 + 2y_+}{1+2y_+ + 3y_+^2}} \;, \quad y_+ = \frac{r_+}{r_c} \;.
\end{equation}
 
\par To discuss the black hole solutions more thoroughly, let us consider the black hole phase space of the RNdS black hole, following Refs. \cite{Bousso1996_Nariai,
AntoniadisBenakli2020_WGCdS}. As we shall see, much of the analysis we perform is based on the metric function Eq. (\ref{eq:f}). To sketch this, we begin with the polynomial, 
\begin{equation}
\Pi(r) \equiv -r^2 f(r) = -r^2 +2Mr -Q^2 + L^2_{dS} r^4 \;.
\end{equation}
\noindent We then determine the discriminant thereof,
{\setlength{\mathindent}{1cm}
\begin{equation} \label{eq:Det}
\Delta \equiv -16 L^{-2}_{dS} \left[27 M^4 L^{-2}_{dS} - M^2(1 + 36Q^2L^{-2}_{dS}) + (Q + 4Q^3L^{-2}_{dS})^2 \right] \;.
\end{equation}}
\noindent To visualise this as a $(M,Q)$ parameter space of solutions, we set $L^2_{dS}=1$, and plot $\Delta = 0$. 
The result is Fig. \ref{fig:sharkfin}, where the boundary corresponds to extremised conditions\footnote{For clarity, we use the term \enquote{extremise} to refer to black holes with degenerate horizons and \enquote{extremal} to refer to the specific case where inner and outer horizons coincide, $r_-=r_+$.}. Pure de Sitter space $(Q=M=0)$ is labeled as Point $O$. Along Line $ON$ lies the Schwarzschild de Sitter family of solutions $(Q=0, M>0),$ with the uncharged Nariai case at point $N$. The Nariai limit corresponds to the upper limit of the black hole mass in de Sitter space-time. The charged Nariai branch of Line $NU$, for which $r_+ \sim r_c$, extends along $NU$, where this represents the upper mass limit of the charged black hole. The opposite branch of Line $OU$ corresponds to \enquote{cold}  black hole solutions, with $r_- = r_+$. On this branch, $Q \sim Q_{ext} \sim r_+$, with $Q_{ext}$ defined in Eq. (\ref{eq:qext}). These branches terminate in the \enquote{ultracold} solution, where $r_- = r_+ = r_c$, and the Hawking temperature goes to zero with the local geometry becoming $\mathbb{M}_2 \times \mathbb{S}^2$.

\par As mentioned earlier, the $Q=M$ line reflects the \enquote{lukewarm} solutions to which black holes in the phase space evolve \cite{Romans1991_ColdLukewarmRNdS,
Brill1993_RNdSextrema,
Mann1995_ChargedBHpairs,
Bousso1999_QuantumStructredS}. With this in mind, we can divide the phase diagram of Fig. \ref{fig:sharkfin} into two regions: the \enquote{colder} $OEU$ region, where $Q>M$, and the warmer $OEN$ region, where $Q<M$. 

\par With this subsection in place, we have established a convenient frame of reference based on the black hole space-time parameters. This provides the necessary formalism needed to address the SCC conjecture in Section \ref{sec:SCCreview}. In our next subsection, we outline the scalar field that we shall use as a proxy for the full non-linear Einstein equations.

\subsection{RNdS perturbations: the quasinormal mode spectrum of the scalar field \label{subsec:RNdSQNF}}\cleqn
\par Let us begin with the generic action for a complex scalar test field charged under $U(1)$, minimally coupled to the Einstein-Hilbert-Maxwell system of Eq. (\ref{eq:action}). The action for this scalar is

\begin{equation}
	S =   \int d^4x \sqrt{-g} \mathcal{L}_{sc.} \;,
\end{equation}
\noindent representing the complex scalar field. The full Lagrangian \cite{HawkingEllis1973_LSSUbook,
BritoCardosoPani2020_Superradiance} is then given by
\begin{equation}
\mathcal{L}_{sc.} = - \frac{1}{2} \left( \mathcal{D}_{\mu} \Phi \right)^{\dagger} \left( \mathcal{D}^{\mu} \Phi \right) - \frac{1}{2}\mu^2 \Phi^{\dagger} \Phi \;,
\end{equation}
with $\mathcal{D}_{\mu} = (\partial_{\mu} - iqA_{\mu})$. The interaction of the charged scalar field with the external electromagnetic field of the black hole $F_{\mu \nu} = \partial_{\mu} A_{\nu} - \partial_{\nu} A_{\mu}$ is introduced through the minimal coupling prescription, in which $\partial_{\mu}$ is replaced by its covariant counterpart, $\mathcal{D}_{\mu}$. Recall that in this stationary black hole context, the only nonzero component of $A_{\mu}$ is $A_t(r)~=~-~Q/r~dt$, the electrostatic four-potential of the black hole.

\par The full, non-linear evolution of the system can be described using the corresponding equations of motion. These separate into the massive, charged Klein-Gordon equation in curved space-time and the Einstein field equations:
\begin{equation}
    \nabla_{\mu} \nabla^{\mu} \Phi - \mu^2 \Phi = 0 \;, \quad G_{\mu \nu} + \Lambda g_{\mu \nu} = 8 \pi G T_{\mu \nu} \;,
\end{equation} 
\noindent for which the stress-energy tensor becomes quadratic in $\Phi$ for this particular model. Higher-order perturbations in the scalar field induce changes in the space-time geometry, as well as in the vector potential \cite{BritoCardosoPani2020_Superradiance}. Following the usual conventions \cite{refBertiCardoso}, we avoid these complications by introducing the linear approximations for the fields $\Phi$ and $g_{\mu \nu}$,
\begin{equation} \label{eq:perts}
    g'_{\mu \nu} = g^{\rm BH}_{\mu \nu} + \delta_{\mu \nu} \;, \quad \Phi'=\Phi^{\rm BG} + \Psi \;.
\end{equation}
\noindent The unperturbed fields, $g^{\rm BH}_{\mu \nu}$ and $\Phi^{\rm BG}$, are referred to as the \enquote{backgrounds}, whilst the \enquote{perturbations}, $\delta_{\mu \nu}$ and $\Psi$, are considered to be very small. If we substitute $g'_{\mu \nu}$ and $\Phi'$ (with $\Phi^{\rm BG}=0$) and linearise the system of equations with respect to $\delta_{\mu \nu}$ and $\Psi$, we find that $\delta_{\mu \nu}$ and $\Psi$ decouple. The metric fluctuations for $\delta_{\mu \nu}$  can then be set to zero and $g^{\rm BH}_{\mu \nu}$ satisfies the vacuum Einstein field equations. In this way, the gravitational sector can be described by the vacuum solution $R_{\mu \nu}=0$ and the backreaction emergent at quadratic order in $\Phi$ is neglected.

\par Let us proceed to the equation of motion for $\Psi$, 
\begin{equation} \label{eq:KGrnds}
\frac{1}{\sqrt{-g}} \left( \partial_{\mu} - iq A_{\mu} \right)\left( \sqrt{-g} g^{\mu \nu} \left( \partial_{\nu} - iq A_{\nu} \right) \Psi \right) = \mu^2 \Psi \;.
\end{equation}
\noindent Recall that we can formulate an ansatz for $\Psi$ derived from the symmetries of the fixed background space-time: in the RNdS case (static, non-rotating, and spherically-symmetric), the wave-function is written in variable-separable form,
\begin{eqnarray} \label{eq:ansatzrnds}
    \Psi_{n \ell } (t,r,\theta,\phi) = \sum_{n=0}^{\infty}\sum_{\ell}^{\infty} \frac{\psi_{n\ell } (r)}{r} \; Y_{\ell } (\theta,\phi) \; e^{-i \omega_{n \ell } t} \;.
\end{eqnarray}
As explained in the introduction, we are concerned only with the $n=0$ \enquote{fundamental mode}, representing the least-damped and thus longest-lived QNM. The angular contribution is expressed using spherical harmonics, for which  $\ell$ represents the angular momentum number. The spherical harmonic function $Y_{\ell } (\theta, \phi )$ satisfies 
\begin{equation}
\nabla^2 Y_{\ell } (\theta, \phi ) = -\frac{\ell (\ell + 1)}{r^2} Y_{\ell } (\theta, \phi) \;.
\end{equation}

\noindent Since the black hole is static, the corresponding ordinary differential equations are time independent. Consequently, the defining QNM behaviour is then fully encapsulated by the radial component. For convenience, we drop the subscripts and write,
{\setlength{\mathindent}{1cm}
\begin{equation}
\frac{d}{dr}\left(r^2 f(r)\frac{d \psi}{dr}\right)+\left(\frac{r^2(\omega+qA_t(r))^2}{f(r)}-\ell(\ell+1)-\mu^{2}r^2 \right) \psi(r)=0\,. \label{eq:radial}
\end{equation}%
}
\noindent Then, redefining $\psi(r)$ as $\psi(r)=\varphi(r)/r$ and employing the tortoise coordinate $r_{\star} = \int dr/f(r)$, we obtain
{\setlength{\mathindent}{1cm}
\begin{equation} \label{eq:odeRNdS}
\frac{d^{2}\varphi(r_{\star})}{dr_{\star}^{2}} + \left[ \omega^2 -V(r) \right] \varphi(r_{\star})=0 \;,
 \end{equation}
\begin{equation}\label{eq:potRNdS}
\mathrm{with} \;\;\; \; V(r)  = f(r) \left[ \frac{\ell (\ell + 1)}{r^2}  +   \frac{f^\prime(r)}{r} + \mu^2 \right] -2\omega q A_t(r)-q^2A_t(r)^2 \;. 
\end{equation}
}
\noindent Note that in the RNdS space-time  $r_{\star}=r_{\star}(r)$ serves as a bijection from $(r_+,r_c)$ to $(-\infty,+\infty)$.

\par Since the RNdS black hole is static and spherically-symmetric, the quasinormal frequency (QNF) eigenvalue problem simplifies to an ordinary differential equation in the radial Schwarzschild coordinate, 
{\setlength{\mathindent}{1cm}
\begin{equation} \label{eq:odefullRNdS}
\frac{d^{2}\varphi(r_{\star})}{dr_{\star}^{2}} + \left[ \left( \omega - \frac{qQ}{r} \right)^2 - f(r) \left[ \frac{\ell (\ell + 1)}{r^2} + \frac{f^{\prime}(r)}{r} + \mu^2 \right]  \right] \varphi(r_{\star})=0 \;,
 \end{equation}}
\noindent where $\ell$ is the angular momentum number and $qQ/r$ represents the coupling between the scalar field charge and the electrostatic four-potential of the black hole $A_t=-Q/r$. The QNM boundary conditions are then imposed,
\begin{equation} \label{eq:BCdS}
\varphi(r_{\star}) \sim 
\cases{e^{-i \left(\omega - \frac{qQ}{r_+} \right)r_{\star}} \;, \quad \quad r \rightarrow r_+ \;\; (r_{\star} \rightarrow - \infty) \;,\\
e^{+i\left(\omega - \frac{qQ}{r_c} \right)r_{\star}} \;, \quad  \quad r \rightarrow r_c \;\; (r_{\star} \rightarrow + \infty) \;,}{}
\end{equation} 
\noindent with radiation purely ingoing at the event horizon and purely outgoing at the de Sitter horizon  \cite{KonoplyaZhidenko2014_RNdSrprc}. 

\par To compute the QNF corresponding to the QNM requires solving for the eigenvalue problem of Eq. (\ref{eq:odefullRNdS}) subjected to the boundary conditions of Eq. (\ref{eq:BCdS}). In this work, we use a semi-classical WKB-based technique described in Ref. \cite{Papantonopoulos2022}. The QNF is expressed as a series expansion $\omega = \sum_{k = -1} \omega_k L^{-k}$, where $L~=~\sqrt{ \ell (\ell + 1)}$. The series expansion is then inserted into
\begin{eqnarray}
\label{eq:BorelQNF}
\omega  & = & \sqrt{V(r^{max}_{\star})-2 i U} \;, \\
U & \equiv & U(V^{(2)},V^{(3)},V^{(4)},V^{(5)},V^{(6)})  \;.
\end{eqnarray}
\noindent The objective of the method is to solve iteratively for the $\omega_k$ coefficients for increasing orders of $k$. $V(r^{max}_{\star})$ corresponds to the peak of the barrier potential of Eq. (\ref{eq:odefullRNdS}), located at 
\begin{eqnarray} 
 r^{max}_{\star} & \approx & r_0 + r_1 L^{-2} + ...\;, \label{eq:rmax} \\ 
 V(r^{max}_{\star})  & \approx & V_0 + V_1 L^{-2} + ... \;, \label{eq:Vmax}
\end{eqnarray}
where subscripts refer to terms in a series expansion around the peak. The numbered superscripts in Eq. (\ref{eq:BorelQNF}) refer to derivatives $V^j$, taken with respect to the tortoise coordinate, such that
{\setlength{\mathindent}{1cm}
\begin{equation} \label{eq:Vj}
V^{j} = \frac{d^j V(r^{max}_{\star})}{dr^j} = f(r) \frac{d}{dr} \left[ f(r) \frac{d}{dr} \left[... \left[f(r) \frac{d V(r)}{dr} \right]... \right] \right]_{r \rightarrow r^{max}_{\star}} \;.
\end{equation}}
\noindent This method is most reliable in the large-$\ell$ regime, and for small values of $Q$ and $q$. Beyond these limitations, the method allows us to maintain the black hole and scalar field parameters as free variables. 

\par The QNF spectrum of a massive, charged scalar field within the RNdS black hole space-time is particularly interesting. Recall that the QNF can be decomposed into its real and imaginary components, where the real part represents the physical oscillation frequency and the imaginary component is related to the damping rate. Upon introducing a non-zero field charge, the QNM reflection symmetry is broken, leaving us with two distinct sets of QNF solutions, here referred to as $\omega_+$ and $\omega_c$. If the QNM $\varphi$ has a charge $q$, then $\varphi^*$ has a charge $-q$. If $\omega = \omega_a + i \omega_b$ is a QNF associated with the QNM $\varphi$, then $-\omega^*= -\omega_a + i \omega_b$ is a QNF associated with the QNM $\varphi^*$.

\par With the necessary GR background provided in this section and in the appendices supplied, the reader is now fully equipped to proceed with the discussion on the SCC conjecture within spherically-symmetric space-times.

\section{Preserving strong cosmic censorship in spherically-symmetric space-times \label{sec:SCCreview}}\cleqn

\par We shall begin our discussion on SCC with a summary of the various formulations of the SCC conjecture used in the mathematical relativity literature. Supplemented by \ref{app:GRdefs} and \ref{app:conform}, this will provide the required nomenclature and necessary formalism to describe the SCC conjecture in a mathematically precise way. Thereafter, we can begin in earnest our discussion of the preservation of SCC, starting with the simple case of the Schwarzschild black hole and building up to the RNdS case. Unless otherwise stated, throughout this discussion we consider the space-time $(\mathcal{M},g)$ to be a 4D time-orientable Lorentzian manifold.

\subsection{Formal definitions for the cosmic censorship conjectures}\cleqn
\par Intuitively, singularities were understood to correspond to \enquote{places} in space-time where curvature \enquote{blows up} or showcases \enquote{pathological behaviour} \cite{Wald1984_GRbook}. To construct a precise singularity theorem proved difficult; the starting point, however, was the understanding that null and time-like geodesic completeness was a minimal requirement for a space-time to be devoid of singularities  \cite{HawkingEllis1973_LSSUbook}. For our purposes, we shall define the singularity with respect to its relationship to the metric, following the statements used in Refs. \cite{Hawking1966_Singularities,HawkingEllis1973_LSSUbook,Penrose1973_CC3},
\begin{proposition}[Singularity]
    Consider $(\mathcal{M},g)$ to be a 4D time-orientable Lorentzian manifold. Then a singularity is denoted by a future-directed future-inextendible time-like curve $C \subset \mathcal{M}$. 
\end{proposition}

\noindent With the singularity defined, we can proceed to the definition of the WCC conjecture,
\setcounter{theorem}{0}

\begin{conjecture}[\textbf{Weak Cosmic Censorship}]
    Consider the strongly causal space-time $(\mathcal{M},g)$ that is asymptotically-flat at null infinity. Then $(\mathcal{M},g)$ contains no naked singularities with respect to $J^-(\mathscr{I^+})$ if and only if $J^-(\mathscr{I^+})$ is globally hyperbolic.\footnote{For additional details on the notation used here, see Table \ref{table:conformalconvention} and  \ref{app:conform}.}
 \end{conjecture}

\par The definition of the SCC conjecture can then be encapsulated by the statement from Ref. \cite{Dafermos:2012np},
\begin{conjecture}[\textbf{Strong Cosmic Censorship}]
For generic vacuum data sets, the maximal future Cauchy development (MFCD) $(\mathcal{M},g_{ab})$ defined in Theorem \ref{thm:MCD} is inextendible as a suitably regular Lorentzian manifold.
\end{conjecture}
\noindent The MFCD is a mathematically-rigorous way in which to describe the evolution of the metric. As we shall see in the following discussion, a single version of the SCC conjecture is too rigid to suit the wide variety of space-times to which the SCC conjecture is applied. Rather, the criterion of extendability is expanded to establish multiple versions of the SCC, with varying relative \enquote{strength}. Using the usual conventions in the literature \cite{Wald1984_GRbook,
HawkingEllis1973_LSSUbook,
Dafermos2018_BlueShiftLambda} and the definitions of smoothness and differentiability introduced in \ref{app:GRdefs}, we list here the different versions of the SCC considered in this work, in order of decreasing strength. We begin with the most satisfactory condition for the SCC conjecture \cite{Christodoulou2008_SCC},
\begin{conjecture}[$C^0$-formulation of Strong Cosmic Censorship] 
For generic vacuum data sets, the MFCD $(\mathcal{M},g_{ab})$ is inextendible past the Cauchy horizon as a $C^0$ manifold.
\end{conjecture}
\noindent This implies that there are no extensions of the metric beyond the Cauchy horizon that preserve the continuity of the metric, without requiring further differentiability. The Cauchy horizon is treated like a singularity. The next strongest statement is attributed to Christodoulou \cite{Christodoulou2008_SCC},
\begin{conjecture}[Christodoulou-formulation of Strong Cosmic Censorship] \label{conj:Chris}
For generic vacuum data sets, the MFCD is inextendible past the Cauchy horizon as a weak solution of the Einstein field equations, such that the Christoffel symbols associated with the $C^0$ metric are locally $L^2$ (i.e. square-integrable when multiplied by any smooth test function of compact support).
\end{conjecture}
\noindent With this reformulation of the SCC conjecture, the metric is not continuous across the Cauchy horizon as a \enquote{weak solution} such that the \enquote{blow up} is in the local Sobolev space i.e. $H^1_{loc}$ (see Definitions \ref{def:weak} and \ref{def:Sobolev} of \ref{app:GRdefs}). 
 In other words, this is the least regular case for which the metric continuously satisfies the Einstein field equations, albeit as a weak solution. Finally, we come to the weakest version of the SCC conjecture considered here,
\begin{conjecture}[$C^2$-formulation of Strong Cosmic Censorship] 
For generic vacuum data sets, the MFCD $(\mathcal{M},g_{ab})$ is inextendible past the Cauchy horizon as a $C^2$ manifold.
\end{conjecture}
\noindent This serves as the lowest threshold utilised when testing the SCC conjecture, thereby serving as a useful test case but not a satisfactory proof for the preservation of SCC.
\par With these definitions in place for the different versions of the SCC conjecture and their relative hierarchy, we can proceed to a study of the conjecture in the context of spherically-symmetric space-times. In so doing, we shall describe how each of these conjectures came to be and how they have been applied to date.

\subsection{SCC in Schwarzschild and RN black holes}\cleqn
\par Recall that for a Schwarzschild black hole space-time, the event horizon is located at $r=2M$ and corresponds to a coordinate singularity. The curvature singularity at $r=0$ is space-like. Considerations of the SCC conjecture in this space-time can be traced back to sources such as Ref. \cite{Gravitation1973}, where an observer falling past the horizon and towards the Schwarzschild singularity was discussed rigorously for the first time, with the observer ripped apart by tidal forces before ever reaching $r=0$. While this \enquote{spaghettification} of the in-falling observer captured the popular imagination as an indication of what black hole interiors could be like, a formal proof that supported the strongest case for SCC remained elusive until recently. As mentioned at the beginning of this note, Sbierski proved that the metric cannot be extended continuously past $r=0$, formally demonstrating that the $C^0$-formulation of SCC is valid in the Schwarzschild black hole \cite{Sbierski:2015nta}.

\par
\begin{figure}[t]
    \centering
    \includegraphics[width=0.7\linewidth]{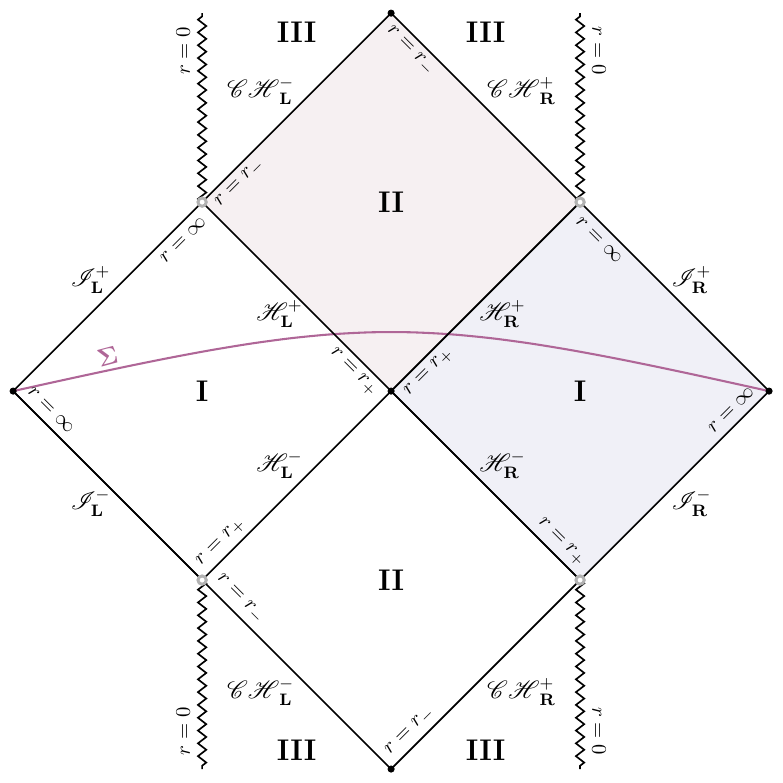}
    \caption{\textit{Conformal diagram for the extended RN black hole space-time, with Cauchy surface $\Sigma$.
    }}
    \label{fig:RN_scc}
\end{figure}

 Let us now move on to the RN black hole space-time, which we sketch in Fig. \ref{fig:RN_scc}. There, the shaded regions refer to the asymptotically-flat universe exterior to the black hole (Region I) and the black hole interior (Region II). The future and past Cauchy horizons are denoted, respectively, by $\mathscr{CH}^+$ and $\mathscr{CH}^-$. Our discussion is concerned only with the \enquote{right} or \enquote{ingoing} horizons, denoted by the subscript \enquote{R}. Recall that the internal structure of the Kerr and the RN black hole space-times are similar, such that arguments corresponding to one can apply also to the other \cite{Landsman2022_PenroseContextualised}. For this reason, we mention results concerning the Kerr space-time here, even though the (axisymmetric) rotating space-time is beyond the scope of this work.

\par For RN black holes (as well as rotating Kerr black holes), the interior is complicated by the presence of the Cauchy horizon at $r = r_-$. For $r < r_-$, surfaces of constant $r$ are time-like such that the curvature singularity at $r=0$ is time-like. The solution is thus time-like geodesically complete but space-like (and null) geodesically incomplete. Consider a space-like surface, $\Sigma$, upon which we prescribe initial data for the asymptotically-flat metric $M$. The metric can then be determined uniquely up to the Cauchy horizon, but not beyond. That is, past $r = r_-$, the evolution of the initial data is governed by unknown boundary conditions at $r=0$. Mathematically, the MFCD of initial data posed on the Cauchy surface is considered incomplete but smoothly extendible beyond the Cauchy horizon. All incomplete geodesics can pass this horizon. For the in-falling observer, this means a safe journey past the Cauchy horizon. For the theory, this corresponds to a failure in determinism. 

\par However, in 1973 Penrose put forth a statement for the Kerr black hole suggesting that for generic asymptotically-flat initial data, the MFCD is inextendible as a continuous Lorentzian metric \cite{Penrose1999_CC}. In other words, this is a claim for the $C^0$-formulation of SCC and a possible way in which SCC could be preserved. If we consider the scalar field $\Phi$ as a linear proxy for the full non-linear Einstein equations, then Penrose's formulation relies on the notion that the scalar field is badly behaved at the Cauchy horizon. For the non-linear case, this corresponds to the energy blowing up at $r=r_-$. The reasoning for this is as follows: Consider two observers external to the RN black hole, a time-like observer $A$ remaining in the exterior region (shaded Region I) and a time-like observer $B$ falling into the black hole (shaded Region II). Observer $A$ will reach future infinity at infinite proper time, observer $B$ will reach the Cauchy horizon at finite proper time. If observer $A$ sends periodic (in $A$'s time) signals to observer $B$, observer $B$ will perceive them as incoming with greater frequency as a result of this gravitational time dilation. For observer $B$, the wavelength of the incoming signal appears to be compressed or \enquote{blue-shifted}. 

\par By this logic, the frequency of an oscillating scalar field $\Phi$ entering the black hole appears to increase infinitely as it reaches the Cauchy horizon. This corresponds to an infinite amplification of the energy of this incoming $\Phi$ field. We can consider the implications of this for the non-linear problem by recalling that this runaway \enquote{blue-shift amplification} of energy will manifest in the Einstein field equations as an infinitely large contribution from the components of the stress-energy tensor. This implies that space-time curvature becomes infinitely large or \enquote{blows up}, which corresponds to the formation of a singularity. 

\par However, Penrose's formulation does not take into account the global black hole metric. It was recently argued by Dafermos \textit{et al.} \cite{Dafermos:2017dbw,Dafermos:2014jwa} that the well-known stability of the Kerr metric exterior in the wake of some space-time perturbation \cite{Teukolsky1973_KerrStability-I,TeukolskyPress1973_KerrStability-II} implies that the incoming scalar field experiences an inverse polynomial decay in time along the event horizon. This suggests that the energy of $\Phi$ at the event horizon is not sufficient to trigger the runaway blue-shift amplification needed to prevent the passage of $\Phi$ across the Cauchy horizon. Decay in the exterior is therefore in competition with the blue-shift mechanism at the Cauchy horizon. The consequence of this argument is that only for smooth localised initial data in sub-extremal Kerr (and, equivalently, RN) black hole space-times, the local energy of $\Phi$ blows up at the Cauchy horizon. Here, the local energy is the integral of the energy as measured by a local observer i.e. the first derivative of $\Phi$ in $L^2$. Formally, $\Phi$ is then inextendible in $H^1_{loc}$ across the Cauchy horizon. Thus, the Christodoulou version of SCC holds in the case of the RN black hole \cite{Luk:2015qja}. Recall that this \enquote{blowing-up} of the energy is a requirement for Christodoulou’s formulation of SCC (see Conjecture \ref{conj:Chris}). As such, this was perceived as a possible means of \enquote{saving} SCC by weakening the $C^0$ requirement further, such that only the MFCD of the metric need be inextendible as a continuous Lorentzian manifold whose Christoffel symbols are locally square-integrable. In other words, we can consider this loosely as a relaxation of the $C^0$-formulation of SCC, where the metric cannot be interpreted even as a weak solution of the Einstein equations past the Cauchy horizon.

\par It is interesting to observe that stability in the metric challenges SCC. For SCC to be preserved, the blue-shift mechanism must win over the exterior decay: the exponential growth of the derivatives at the Cauchy horizon must exceed the inverse polynomial decay of $\Phi$. To ensure this, the decay of $\Phi$ must not be too fast as to overwhelm the blue-shift. We shall see that it is this requirement that leads us to the works of Refs. \cite{Hintz2015_SCC,Costa2016_SCC}, where QNMs dominate the exponential behaviour for $\Lambda >0$. 

\par However, to finalise our discussion on the current state of the SCC, we note that as discussed by Franzen \cite{Franzen:2014sqa}, the blue-shift may not be enough to rescue SCC if we consider also the amplitude of $\Phi$. This is because the blue-shifting (the increasing frequency of the oscillations entering the black hole) affects the derivatives of $\Phi$ but has no influence on its amplitude. Under this argument, it may be shown that $\Phi$ remains uniformly bounded on the black hole interior for the sub-extremal RN black hole and extends continuously past the Cauchy horizon, violating the $C^0$ formulation of SCC.

\par Let us now proceed to the case of the RNdS black hole space-time, where we shall see the effect of the positive cosmological constant on the preservation of SCC. Specifically, we shall discuss the relationship between the \enquote{blue-shift} (i.e. the mechanism responsible for the exponential growth of the perturbations) affecting the derivatives of $\Phi$ in the interior, and a competing \enquote{red-shift} effect emergent in the case of de Sitter space-times. This is constructed in analogy to \enquote{Price's Law} in asymptotically-flat space-time \cite{Dafermos:2003yw}, where a \enquote{red-shift} corresponds to a shift of late-stage radiation to spatial infinity. In asymptotically-de Sitter space-time, if an observer $A$ crossing $\mathscr{H}^+_R$ emits a signal at a constant rate with respect to their own proper time, the frequency of the signal as received by an observer $B$ crossing $\mathscr{H}^+_R$ at a later time is shifted to $\mathscr{C}^+_R$.

\subsection{SCC in RNdS black holes \label{sec:SCCRNdSrev}}\cleqn
\par
\begin{figure}[t]
    \centering
    \includegraphics[width=0.6\linewidth]{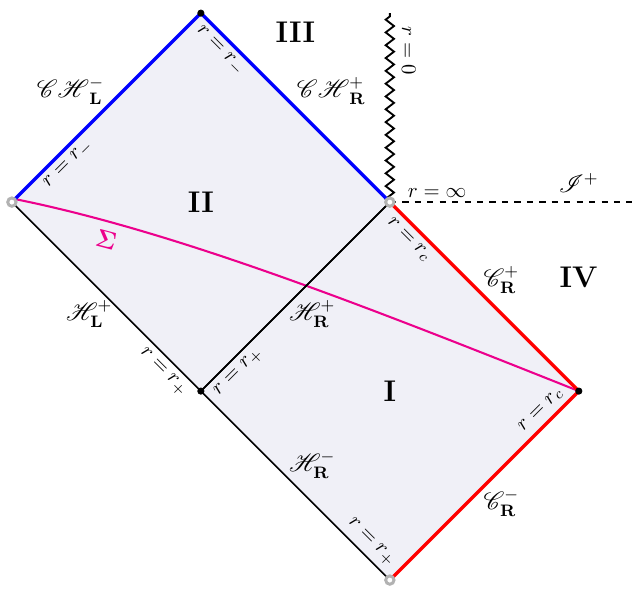}
    \caption{\textit{From the RNdS Penrose diagram of Fig. \ref{fig:Penrose_RNdS}, we consider the competing blue-shift and red-shift mechanisms.}}
    \label{fig:RNdS_scc}
\end{figure}

 Finally, let us turn to the case of the RNdS black hole. Once again, the stability of the metric in the shaded Region I of Fig. \ref{fig:RNdS_scc} (for which $r_+ < r < r_c$) has been proven (see Ref. \cite{refIKchap6}). However, the decay of $\Phi$ differs: while generic initial data defined on the Cauchy surface decays inverse polynomially on the event horizon in asymptotically-flat space-time, the decay becomes exponential in asymptotically-de Sitter space-time \cite{Dafermos:2007jd}. Intuitively, we understand this to be a consequence of $\Lambda >0$: an accelerating expansion of the universe and the associated cosmological horizon leads to a \enquote{red-shifting} effect. There is competition between the gravitational pull of the black hole and the repulsive energy of the cosmological constant. That is, between the event horizon and the cosmological horizon lies a potential well that influences the decay rate of the scalar perturbations. As a consequence of this, $\Phi$ at the black hole exterior then experiences a red-shifted wavelength and a decrease in energy, corresponding to exponential damping. The asymptotics of $\Phi$ become governed by QNM behaviour.

\par This faster decay rate must not overcome the blue-shifting at the Cauchy horizon if SCC is to be preserved. To quantify the exponential rates precisely, one must take into account the influence of all three physical horizons. This problem was recently resolved through Refs. \cite{Hintz2015_SCC}, which focused on the behaviour of the linear perturbations of smooth initial data. This analytical means of scaling the competing blue- and red-shifting phenomena makes use of the exponential dependence on $\kappa_-$ at the Cauchy horizon and the least-damped QNF. 

\par The main finding of Ref. \cite{Hintz2015_SCC} is as follows, For a non-degenerate RNdS black hole of dimension $d \geq 4$, there exists some $\beta>0$ dependent only on the black hole parameters such that the exponential decay for massive and neutral scalar fields is governed by the expression, 
\begin{equation} \label{eq:Hintz}
\vert \Phi  \vert \leq C e^{-\beta  t} \;, \quad \beta \equiv -\frac{\mathfrak{I}m \{ \omega^{n=0} \}}{\vert \kappa_- \vert} \;,
\end{equation}  
\noindent for small mass $m>0$. As before, $\Phi$ is a linear scalar perturbation, $C\geq 0$ is a constant and $\mathfrak{I}m \{ \omega^{n=0} \}$ is the least-damped QNF. 
 
\par In this case, $\Phi$ becomes continuous up to the Cauchy horizon, and the derivatives of $\Phi$ lie in the Sobolev space $H^{ \left(\beta + \frac{1}{2} \right) - \epsilon} \; \forall \epsilon > 0$. To satisfy the $C^2$ formulation of the SCC, $\beta<1$. However, the stronger Christodoulou formulation requires that $\beta<1/2$ for $\Phi$ to be inextendible across $\mathscr{CH}^+_R$ in $H^1_{loc}$ for SCC to be preserved \cite{Hintz2015_SCC,Costa2016_SCC}.
 
\par As such, the QNM study of SCC preservation that we shall perform in Section \ref{subsec:SCCinRNdS} will be predicated on the condition,
\begin{equation} \label{eq:savingSCC}
\beta \equiv -\frac{\mathfrak{I}m \{ \omega^{n=0} \}}{\vert \kappa_- \vert} < \frac{1}{2} \;.
\end{equation}
\noindent In particular, we shall investigate where in the RNdS phase space we are more likely to observe a violation of the Christodoulou formulation of SCC. Certain examples have already been noted in the literature \cite{Cardoso2018_QNMsSCC,DiasEperonReallSantos2018_SCC,DiasReallSantos2018_SCC_Rough}. 

\par This indicates that $\Lambda$ can cause a failure of SCC. However, we add as a final comment on the SCC that if we accept a relaxation on the requirement of the smoothness of the initial data, the SCC can remain valid. Specifically, to rescue the Christodoulou formulation of SCC, Dafermos \textit{et al.} proved in Ref. \cite{Dafermos2018_BlueShiftLambda} that by relaxing the requirement on the regularity class $-$ considering initial data that is in the class $H_{loc}^1 \times L^2_{loc}$ $-$ the local energy does indeed blow up at the Cauchy horizon for sub-extremal RNdS black holes. That the preservation of SCC depends on the use of rough initial data is further corroborated in Ref. \cite{DiasReallSantos2018_SCC_Rough}.

\section{Violations of SCC in RNdS black holes \label{subsec:SCCinRNdS}}\cleqn
\begin{figure}[t]
        \centering
        \includegraphics[width=0.6\textwidth]{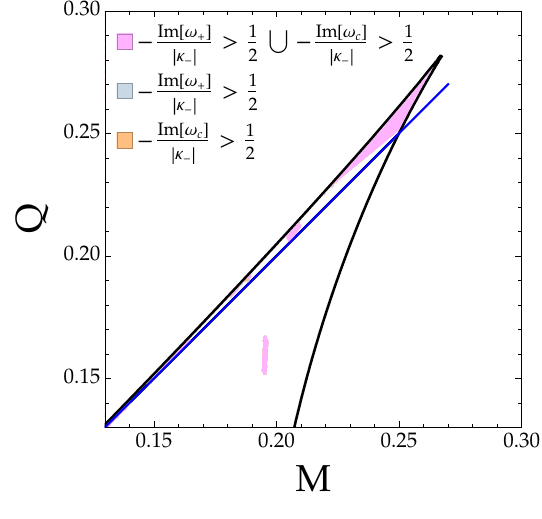}
        \caption{\textit{With $L^2_{dS}=1$ and for $\mu=\ell=1$ and $q=0.1$, we shade the parameter space in which $\omega_+$ (blue), $\omega_c$ (orange), and both families (magenta) violate the condition for SCC preservation. \label{fig:SCCell1q01Mu1}}}
\end{figure}

\begin{figure}[t]
        \centering
       \includegraphics[width=.6\textwidth]{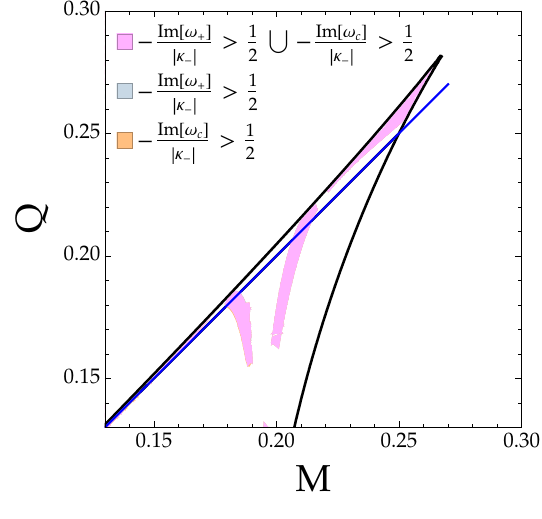}
        \caption{\textit{With $L^2_{dS}=1$ and for $\mu=0.1$, $q=1$, and $\ell=1$, we shade the parameter space in which $\omega_+$ (blue), $\omega_c$ (orange), and both families (magenta) violate the condition for SCC preservation.\label{fig:SCCell1Mu01q1}}}
\end{figure}

\par As discussed, there is an expectation based on Penrose's SCC conjecture that the presence of a Cauchy horizon in the RN back hole interior leads to the infinite amplification of perturbations in the interior region through a \enquote{blue-shift} mechanism. The Cauchy horizon then is destabilised and behaves as a singularity beyond which the initial data cannot be extended. For the RNdS case, the presence of the cosmological horizon leads to an exponential decay of perturbations entering from the black hole event horizon. Without sufficient energy to trigger the runaway blue-shift amplification upon which SCC preservation depends, it may be possible for $\Phi$ to cross the Cauchy horizon and thereby violate determinism in the process.

\par As discussed in Section \ref{sec:SCCRNdSrev}, we can set an upper bound on the damping rate of the incoming scalar field $\Phi$ above which the red-shift effect will overcome the blue-shift. The criterion for the preservation of (the Christodoulou formulation of) SCC is provided in Eqs. (\ref{eq:Hintz}) and (\ref{eq:savingSCC}). This provides us with a means by which to test whether the QNF spectra that we calculate for a variety of black hole and scalar field parameters satisfy the condition for SCC preservation. 
\par Using the semi-classical technique introduced in Section \ref{subsec:RNdSQNF} and described in detail in Ref. \cite{Papantonopoulos2022}, we explore the RNdS $(M,Q)$ parameter space provided in Fig. \ref{fig:sharkfin} in order to determine for which space-time and scalar field parameters we find evidence for $\beta >1/2$ for $\ell =1$. 

\par We find that the SCC is largely preserved within the sharkfin; we only find evidence of its violation for intermediate and large values of $M$ and $Q$. In particular, we consistently find evidence that SCC is violated within the shaded $OEU$ region of the sharkfin and on certain points on the $OU$ line, particularly near $M \sim Q \sim 0.089 $ for near-zero mass and charge. {\color{black}{For $\ell=1$ and $q=0.1$, we find that $\mu=0.1$ and $\mu=1$ violate identical regions, yielding Fig. \ref{fig:SCCell1q01Mu1}. When $q=1$, however, a larger region of the parameter space is violated, extending from the extremal $r_- \sim r_+$ regime.}} 

\par
\begin{figure}[t]
\centering
\includegraphics[width=.6\textwidth]{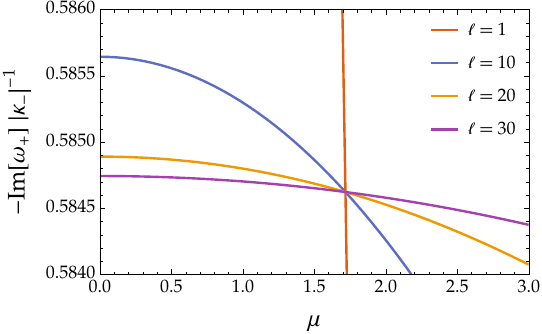}
\caption{\textit{For $L^2_{dS}=1$, $M=0.157$, and $Q=0.158$, we plot the critical mass $\mu_{crit} \sim 1.7$ corresponding to a violation of the condition for SCC preservation.}
\label{fig:mucritSCC}}
\end{figure}

 Finally, we note with interest that for a very small parameter space on the $OU$ line corresponding to extremal black holes, we observe in Fig. \ref{fig:mucritSCC} a \enquote{critical mass} $\mu_{crit}$ for which $\beta > 1/2$. Note that this critical mass value signifies a specific scalar field mass value below which we observe an inverted relationship between the QNM damping rate and the angular momentum number. While we expect $\vert \mathcal{I}m \{ \omega \} \vert$ to increase with $\ell$, for $\mu < \mu_{crit}$, the imaginary part decreases with increasing $\ell$ (as first observed, to our knowledge, in Ref. \cite{Lagos2020_Anomalous} for Schwarzschild and Kerr black holes).  

\section{Conclusions}\cleqn

\par In this note, we have provided a pedagogical treatment of the SCC conjecture in spherically-symmetric black holes, culminating in the RNdS black hole space-time. We have included an extensive review of the SCC conjecture and its various formulations in the hopes of providing an accessible overview of the topic for members of the community interested in the problem but not well-versed in the mathematical relativity literature. 

\par Our discussion on the SCC conjecture also facilitated an exploration of the causal structure of the RNdS black hole, and the  parameter space of RNdS black hole solutions. This allowed us to draw attention to the antithetical role of horizon stability in the preservation of WCC and SCC: where singularities exist, the WCC conjecture requires the existence and stability of event horizons. The SCC conjecture, on the other hand, requires the absence or instability of Cauchy horizons in all cases. A similar observation was made in Ref. \cite{Landsman2022_PenroseContextualised}.

\par  While the notion of determinism and its preservation within GR is a fascinating topic in and of itself, this highlights in particular the importance of understanding QNMs within in the RNdS context. While stability for the charged and massive scalar field is well-established for $\ell >0$ \cite{KonoplyaZhidenko2014_RNdSrprc}, it is interesting that perturbations at the black hole exterior can be used to infer information about the the dynamics of the scalar field within the black hole interior.

\par From our semi-classical analysis of the QNF spectrum within a RNdS background, we note that the condition for the preservation of SCC was violated near $M \sim Q \sim 0.089 $. For non-zero $\mu$, black holes \enquote{colder} than the cosmological horizon were associated with SCC violations. In fact, this question of whether large, cold black holes respect cosmic censorship has recently been investigated in Ref. \cite{Hod2023_ColdSCC}, where it was found that near-extremal charged black holes may collapse into naked singularities during photon emission. This certainly invites further study, particularly concerning the possible constraints we can consequentially apply to the black hole and the photon emitted, as well as the possible implications this may have for SCC preservation.

\subsection*{Acknowledgements}
AC acknowledges the support of the National Research Foundation (NRF) of South Africa and Department of Science and Innovation through the SA-CERN programme, as well as 
that of a Campus France scholarship and a research grant from the L’Oréal-UNESCO's \textit{For Women in Science} Programme. AC is currently supported by the Initiative Physique des Infinis (IPI), a research training program of the Idex SUPER at Sorbonne Universit\'e.
ASC is supported in part by the NRF. SCP is supported by the National Research Foundation of Korea (NRF) grant funded by the Korea government (MSIT) RS-2023-00283129 and RS-2024-00340153. The authors thank Hajar Noshad for discussions and collaboration during the initial stages of this project. We also thank Roman Konoplya for his indispensable comments and advice.

\appendix
\section{Mathematical preliminaries on cosmic censorship\label{app:GRdefs}}\cleqn
\renewcommand{\theequation}{A.\arabic{equation}}
\renewcommand{\thetheorem}{A.\arabic{theorem}}
\setlength{\parindent}{1em}

\par Recall that a well-defined theory is predictable and deterministic. This innocuous statement, often an implicit assumption in our scientific thinking, has been the source of ongoing debate within the spheres of physics and philosophy alike. Through the Laplacian view of the universe \cite{Laplace1902_Probabilities}, Werndl argues in Ref. \cite{TimpeGriffithLevy2017_FreeWill} that determinism is an ontological position while predictability is an epistemological one. In other words, determinism is an inherent quality of the state, while predictability is our description of the state which is confirmed retrospectively by our observations of the evolution of the state. Determinism is the requirement that it evolves uniquely: the properties of a state are a direct consequence of its past and serve as the cause of its future. This can be considered a fundamental axiom in quantum mechanics, for example. Predictability, on the other hand, is a close cousin of reproducibility: it is the requirement that this unique evolution can be accurately described by the theory in place. Chaotic systems are then fine examples of deterministic models that are unpredictable.

\par Within the context of GR, the integrity of determinism is called into question when we consider a black hole interior within which a Cauchy horizon can be found  \cite{Dafermos2003_RNinterior,Smeenk2021_DeterminismGR}. It is this question of determinism that serves as the underlying principle of the SCC conjecture. 
Particularly among mathematicians, the argument 
concerns the evolution of the metric over the Cauchy horizon and the regularity conditions that must be satisfied for the corresponding equations of motion to make sense. To understand the evolution of these partial differential equations and how regularity conditions influence the preservation of SCC, the 
 scalar field serves as a convenient proxy for the full non-linear Einstein field equations. 
 However, to venture into these formal arguments, there are a number of mathematical preliminaries that must be specified first. To address these, we rely primarily on Refs. \cite{Lawrence2022_PDEs,HawkingEllis1973_LSSUbook,Wald1984_GRbook}. 

\par Furthermore, when discussing the evolution of partial differential equations and the physical insights we might glean from them, it is imperative that we do so in the appropriate linear space. Note also that the principles of functional analysis usually employed for this purpose of interpreting the partial differential equation problem can only be applied after the problem of interest has been restructured in the form,
\begin{equation}
A: X \rightarrow Y \;,
\end{equation}
\noindent where $A$ serves as the operator encoding the structure, boundary conditions, etc. of the partial differential equation and $X,Y$ are the spaces of functions. As we shall see, the appropriate space for the linear partial differential equations we consider here is the \enquote{Sobolev space}.

\par To introduce our notations and conventions, let us begin with the familiar properties of smoothness and continuity. Generally, the \enquote{smoothness} of a function is a property related to the highest order of its continuous derivatives. For a function $f$ defined on an open set on the real line $\mathbb{R}$ and $k \in \mathbb{Z}^{+}$, recall the definition for the differentiability class of a function: 
\begin{definition}[Differentiability class] \label{def:diffclass}
     A function $f \mathbb{:} \mathcal{M} \subset \mathbb{R}^n \rightarrow \mathbb{R}$ belongs to differentiability class $C^k$ if it is $k$-differentiable on $\mathcal{M}$ and its $k$-order gradient is continuous on $\mathcal{M}$.
\end{definition}
\noindent As such, $f$ is considered to be in the class $C^{\infty}$, ``smooth" or ``infinitely differentiable", if its derivatives to all orders exist and are continuous. For example, a $C^{\infty}$ function such as $f(x) = e^{2x}$ has continuous derivatives at all orders. Formally, we can then define \enquote{smoothness}: 

\begin{definition}[Smoothness]
    A smooth curve $C$ on a manifold $\mathcal{M}$ is a $\mathcal{C}^{\infty}$ map of $\mathbb{R}$ into $\mathcal{M}$, $C \mathbb{:R} \rightarrow \mathcal{M}.$ At each point $p \in \mathcal{M}$ lying on the curve $C$, we can associate with $C$ a tangent vector $T \in V_p$. Then the derivative of the function $f \circ C \mathbb{:R} \rightarrow \mathbb{R}$ evaluated at $p$ is $T(f) = d(f \circ C)/dt$.
\end{definition}
\noindent A function that does not meet this criterion is then considered \enquote{rough}. 

\par We can see that there is a hierarchy of strictness in place, where higher values of $k$ correspond to greater strictness. Recall also that differentiability is a stronger condition than continuity. For our purposes, the preservation of SCC hinges on the inextendability of the metric across the horizon: the strongest version of SCC therefore corresponds to a $C^0$ case, where the metric cannot be extended continuously across the horizon. As mentioned in Section \ref{sec:SCCreview}, this condition is not satisfied for the RN black hole's Cauchy horizon \cite{Christodoulou2008_SCC,Dafermos2003_RNinterior}. 

\par In order to allow for a slight relaxation of the condition for differentiability, it becomes necessary to introduce the \enquote{weak derivative}, which we shall refer to simply as a generalisation of the derivative defined on the space for functions that are integrable but not assumed to be differentiable. For notational convenience, let us use $C^{\infty}_c (U)$ to denote the space of infinitely differentiable functions $\varphi \mathbb{:}\;U \rightarrow \mathbb{R}$. Note that the $c$ subscript refers to compact support in $U$, where the term \enquote{compact support} conveys that the function is defined only on the compact space $U$ and is zero elsewhere.

\begin{definition}[Weak partial derivative\label{def:weak}]
    Let us suppose that $u$ and $v$ are locally summable on the Lebesgue space $L^1$ (i.e. $u, v \in L^1_{loc}(U)$). Suppose also that $\alpha$ is a multi-index such that $\alpha = (\alpha_1,...,\alpha_n)$ and $\vert \alpha \vert = \alpha_1 + ... + \alpha_n$. Then we can say that $v$ is the $\alpha^{\mathrm{th}}$-weak partial derivative of $u$,
\begin{equation}
D^{\alpha} u = v \;,
\end{equation} 
\noindent provided that for all test functions $\varphi \in C_c^{\infty}(U)$,
\begin{equation}
\int_U dx \; u D^{\alpha} \varphi  = (-1)^{\vert \alpha \vert} \int_U dx \; v \varphi  \;.
\end{equation}
\end{definition}
\noindent The \enquote{strong} solution to a (partial) differential equation is expected to be smooth, and to hold point-wise for every point in $U$. The \enquote{weak} solution, on the other hand, satisfies the equation without meeting the same strict criteria (e.g. the solution can only be integrable rather than continuously differentiable). In this way, we generalise or broaden the class of functions that can serve as solutions to the equation.

\par It becomes useful to define a space in which these weak (partial) derivatives are well-defined. This is the \enquote{Sobolev} space \cite{Lawrence2022_PDEs}:
\begin{definition}[Sobolev space\label{def:Sobolev}]
    Let us consider the open set $U \subset \mathbb{R}^n$, the Sobolev space $W^{k,p} (U)$ then consists of all locally summable functions $u \mathbb{:}\; U \rightarrow \mathbb{R}$ such that for each multi-index $\vert \alpha \vert  \leq k$, the weak derivative $D^{\alpha}u$ exists and belongs to the space $L^p (U)$ for a fixed $1 \leq p \leq \infty$ and $k \in \mathbb{Z}^+$.
\end{definition}
\noindent Upon introducing the norm to $W^{k,p} (U)$,
\begin{equation} \label{eq:Banach}
    \vert \vert u \vert \vert_{W^{k,p} (U)} \equiv \sum_{\vert \alpha \vert \leq k} \vert \vert D^{\alpha}u \vert \vert_{_{L^p (U)}} \;,
\end{equation}
and $W^{k,p} (U)$ becomes complete and Eq. (\ref{eq:Banach}) represents a Banach space. If we impose $p=2$, $k \in \mathbb{N}_0$, and $U = \mathbb{R}^n$, 
\begin{equation}
H^k (U) = W^{k,2} (U) \;;
\end{equation}
\noindent $H^k (U)$ is then a Hilbert space as well as a Banach space, and $H^0 (U) = L^2 (U)$. Throughout this note, we discuss functions belonging to the space $H^1_{loc} (U)$. This implies that both the function and its first weak derivatives are square-integrable (i.e. they are in $L^2(U)$). The subscript $loc$ conveys that the function need not be globally regular, but retains local regularity properties.

\par The Sobolev space is ideal for defining (less regular) functions and their weak derivatives. As in Definition \ref{def:diffclass}, $k$ relays how differentiable a function is, whilst $p$ tells us about a function's integrability. Spaces with larger $k$ generally contain smoother functions, whereas larger $p$ implies more \enquote{roughness}. For many cases, there is a trade-off between smoothness and integrability:  a function that belongs to $W^{k,p} (\mathbb{R}^n)$ belongs to certain $W^{\ell,q} (\mathbb{R}^n)$ where $\ell < k$ and $p > q$. For a sufficiently large $k$ and $p$, the function can be classified as classically differentiable.

\par These facilitate the formal definitions of the cosmic censorship conjectures, allowing us to make precise our discussion on the nature of SCC and its preservation in various space-time configurations. Since we rely on the initial-value approach to GR to define the cosmic censorship conjectures, we close this appendix with the introduction of the concept of the \enquote{maximal future Cauchy development} (MFCD), through the theorem of Choquet-Bruhat and Geroch (see Chapters 7.6 of Ref. \cite{HawkingEllis1973_LSSUbook} and 10.2 of Ref. \cite{Wald1984_GRbook}). While we make use of a scalar field as a proxy for the full non-linear problem, it is the MFCD that is considered in the rigorous formal analysis of SCC preservation. 
\begin{theorem}[Maximal future Cauchy development] \label{thm:MCD}
Let $\Sigma$ be a three-\\ -dimensional $C^{\infty}$ manifold, let $h_{ab}$ be a smooth Riemannian metric on $\Sigma$, and let $K_{ab}$ be a smooth symmetric tensor field on $\Sigma$. Suppose that the initial data $(\Sigma,h_{ab},K_{ab})$ satisfies the vacuum constraint equations,
\begin{eqnarray}
0 & = & D_b K^b_{\;\;a} - D_a K^b_{\;\;b} \;,\\
0 & = & \frac{1}{2} \left(^{(3)}R + (K^a_{\;\;a})^2 -K_{ab}K^{ab} \right) \;,
\end{eqnarray}
\noindent where $D_a$ is the derivative operator associated with the metric $h_{ab}$. Then there exists a unique $C^{\infty}$ space-time $(\mathcal{M}, g_{ab})$ known as the maximal future Cauchy development of $(\Sigma,h_{ab},K_{ab})$ that satisfies the following properties:
\begin{itemize}
\item[$(i)$] $(\mathcal{M}, g_{ab})$ is a solution to the Einstein field equations.
\item[$(ii)$] $(\mathcal{M}, g_{ab})$ is globally hyperbolic\footnote{A set $\mathcal{N}$ is said to be globally hyperbolic if for any two points $p,q \in \mathcal{N},$ $J^+ (p) \cap J^-(q)$ is compact and contained in $\mathcal{N}$. In other words, $J^+ (p) \cap J^-(q)$ does not contain any points on the edge of space-time i.e. at spatial infinity or at a singularity \cite{HawkingEllis1973_LSSUbook}.} with a Cauchy surface\footnote{A Cauchy surface $\Sigma$ in $\mathcal{M},g_{ab})$ is a surface that is intersected exactly once by every inextendible null and time-like curve in $(\mathcal{M},g_{ab})$ \cite{Penrose1964_CC1}.} $\Sigma$.
\item[$(iii)$] $h_{ab}$ and $K_{ab}$ are the induced metric and extrinsic curvature, respectively, of $\Sigma$.
\item[$(iv)$] A space-time $(\mathcal{M}^{\prime}, g^{\prime}_{ab})$ satisfying properties $(i)-(iii)$ can be mapped isometerically into a subset of $(\mathcal{M}, g_{ab})$. To elaborate, suppose $(\Sigma,h_{ab},K_{ab})$ and $(\Sigma^{\prime},h^{\prime}_{ab},K^{\prime}_{ab})$ are initial data sets with maximal developments $(\mathcal{M}, g_{ab})$ and $(\mathcal{M}^{\prime}, g^{\prime}_{ab})$, respectively. Suppose also that there is a diffeomorphism between $S \subset \Sigma$ and $S^{\prime} \subset \Sigma^{\prime}$ that carries $(h_{ab},K_{ab})$ on $S$ into $(h^{\prime}_{ab},K^{\prime}_{ab})$ on $S^{\prime}$. Then $D(S)$ in $(\mathcal{M}, g_{ab})$ is isometric to $D(S^{\prime})$ in $(\mathcal{M}^{\prime}, g^{\prime}_{ab})$. Finally, the solution $g_{ab}$ on $\mathcal{M}$ depends continuously on the initial data $(h_{ab},K_{ab})$ on $\Sigma.$
\end{itemize}
\end{theorem}

\noindent 
\section{Horizons and conformal diagrams: the Schwarzschild example \label{app:conform}}\cleqn
\renewcommand{\theequation}{B.\arabic{equation}}
\renewcommand{\thetheorem}{B.\arabic{theorem}}
\setlength{\parindent}{1em}

\par A black hole is distinguished from other compact bodies by two features: its event horizon and the singularity it encloses \cite{Nature_BlackHoleDefinitions}. Within GR, we consider the black hole through the (semi-)classical lens, such that infalling matter that passes the event horizon becomes trapped within the horizon but matter therein has no possibility of escape.\footnote{Note, however, that the event horizon is not a \enquote{trapped surface}: a closed, space-like 2-surface whose area decreases locally along any future direction, even along outgoing null (light-like) geodesics \cite{Penrose1964_CC1}. For stationary black holes, the event horizon is considered as the boundary of the region \textit{containing} the trapped surface.} Since the causal relationship between the black hole interior $r < r_+$ and the exterior space-time $r>r_+$ proves fundamental to the physics herein pursued, we set aside this appendix to contextualise some of the terminology used. To do so, we largely follow Refs. \cite{HawkingEllis1973_LSSUbook,Wald1984_GRbook,
Wald1995_BHthermQFTbook,Landsman2022_PenroseContextualised} and refer to the simple case of a Schwarzschild black hole in asymptotically flat space-time when necessary.

\par We may begin with a formal definition of the black hole in an isolated system, $viz.$:
\begin{definition}[Black hole]  \label{thm:BH}
Consider a causal and asymptotically flat space-time $(\mathcal{M},g)$. The space-time is said to contain a black hole if $\mathcal{M}$ is not contained in $J^- (\mathscr{I}^+)$. Then the black hole region is defined as $\mathcal{B} \equiv \mathcal{M} - J^-(\mathscr{I^+})$. The event horizon is the boundary of $\mathcal{B}$ in $\mathcal{M}$, $\mathscr{H} \equiv  J^-(\mathscr{I^+}) \cap \mathcal{M}$.
\end{definition}
\noindent Here, $\mathscr{I^+}$ refers to the future null infinity (see Table \ref{table:conformalconvention}) and $J^-$ represents the chronological past. As such, the event horizon is the boundary of the past of $\mathscr{I}^+$, where it is a null hypersurface comprised of future inextendible null geodesics without caustics. 

\par We refer to this black hole as \enquote{stationary}, with $\mathcal{B}$ stationary, if there exists a one-parameter group of isometries\footnote{Recall: an isometry $i: \mathcal{M} \rightarrow \mathcal{M}$ on a space-time $(\mathcal{M},g)$ is a diffeomorphism that leaves the metric invariant.} on the space-time $(\mathcal{M}, g_{ab})$ whose orbits are time-like. These isometries are generated by a \enquote{Killing (vector) field} $\xi^a$ that is unit time-like (i.e. $\xi^a\xi_a \rightarrow -1$) at infinity. A null surface $\mathscr{K}$ to which the Killing field $\xi^a$ is normal is referred to as a \enquote{Killing horizon}. By the \enquote{rigidity theorems} attributed to Carter \cite{Carter1973_Rigidity} and Hawking \cite{HawkingEllis1973_LSSUbook}, the event horizon of a stationary black hole must be a Killing
horizon. 

\par Since $\xi^a\xi_a=0$ on $\mathscr{K}$, the vector $\nabla^b (\xi^a \xi_a)$ is also normal to $\mathscr{K}$, by definition. These vectors must then be proportional at every point on the surface $\mathscr{K}$. As such, we can introduce the \enquote{surface gravity} $\kappa$ through the expression,
\begin{equation}
\nabla^b (\xi^a \xi_a) = -2 \kappa \xi^b \;,
\label{eq:kappadef}
\end{equation}
\noindent where $\kappa$ is constant along each generator \cite{Carter1973_Rigidity,HawkingEllis1973_LSSUbook}. It can be shown \cite{Wald1995_BHthermQFTbook,Wald1984_GRbook} that
\begin{equation}
\kappa = \lim_{r \rightarrow r_+}  a(- \xi^a \xi_a)^{1/2} \;,
\end{equation}
\noindent where $a$ is the magnitude of the acceleration of the orbits of (time-like)
$\xi^a$ in the region outside $\mathscr{K}$. The use of the term \enquote{surface gravity} for $\kappa$ arises from the interpretation of $(- \xi^a \xi_a)^{1/2}$ as a \enquote{gravitational redshift factor}, such that $\kappa$ serves as the 
\enquote{redshifted proper acceleration} of the orbits of $\xi^a$ near the horizon.

\par
\begin{figure}
\centering
        \includegraphics[width=0.95\textwidth]{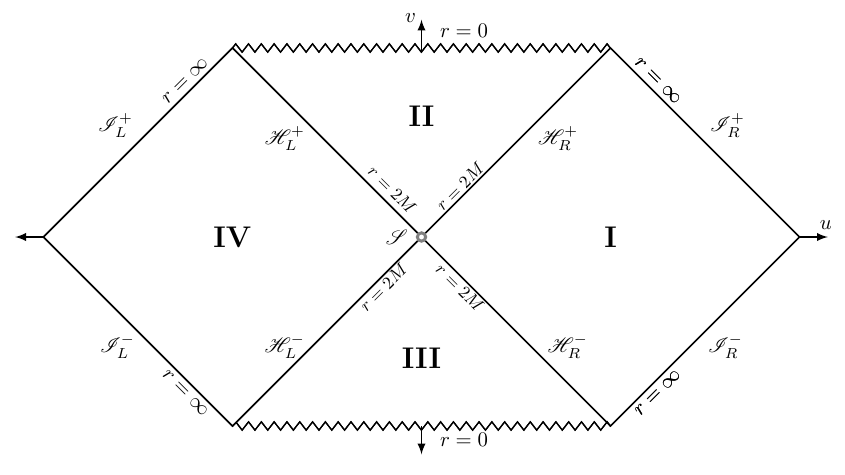}
        \caption{\textit{Conformal diagram for the maximally-extended Schwarzschild space-time.}\label{fig:SchwarzConform}}
\end{figure}

 Finally, we introduce the \enquote{bifurcate Killing horizon}: a pair of null surfaces $\mathscr{K}_A$ and $\mathscr{K}_B$ that intersect on a space-like 2-surface $\mathscr{S}$, the \enquote{bifurcation surface}, such that $\mathscr{K}_A$ and $\mathscr{K}_B$ are each Killing horizons with respect to the same Killing field $\xi^a$. It then follows that $\xi^a$ must vanish on $\mathscr{S}$. Assuming a non-zero $\kappa$, the event horizon of a \enquote{maximally-extended} black hole comprises a branch of a bifurcate Killing horizon (see, for example $\mathscr{H}^+_R$ and $\mathscr{H}^-_R$ of Figure \ref{fig:SchwarzConform}) \cite{Wald1995_BHthermQFTbook}. The \enquote{extremal} black holes for which $\kappa_+=0$, on the other hand, do not possess this bifurcate event horizon.

\par Now that we have clarified our approach to black holes and their horizons, let us proceed to a discussion on the causal relationships between the regions of space-time they establish. A key visual aide in this is the \enquote{conformal space-time diagram}, also known as \enquote{Carter-Penrose diagrams}. These were introduced by Brandon Carter and Roger Penrose to illustrate the structure of infinitely-large black hole space-times. These two dimensional projections are particularly useful as a concise geometric representation of space-time regions and the causal relationships between them, with key features such as event horizons and singularities clearly delineated. 
Conformal diagrams allow us to visualise complex field behaviours within curved space-times. In this section, we provide a brief overview of the construction of these space-time diagrams, using the maximally extended Schwarzschild solution as an example and following Refs. \cite{refHorowitzChap1,Wald1984_GRbook,
HawkingEllis1973_LSSUbook}. 

\par Without loss of information, the conformal diagram must capture the characteristics of the $(3+1)$-dimensional space-time with a $(1+1)$-dimensional diagram. In the case of spherically-symmetric black holes, we are able to suppress the angular dimensions $(\theta,\phi)$ while maintaining the integrity of the global and causal structure of the space-time. To do so, a coordinate transformation must be introduced such that:
\begin{spacing}{0.7}
\begin{itemize}
\item[$\bullet$] radial light rays are always at $\pm 45^{\circ}$ from the vertical axis (recall that $c=1$);
\item[$\bullet$] points in the infinite past or future lie at a finite coordinate distance, at the boundary of the diagram.
\end{itemize}
\end{spacing}
\noindent In other words, the infinite space-time must be transformed into a finite diagram, with angles preserved but distances compactified. This is possible through a conformal rescaling of the metric,
\begin{equation}
g_{\mu \nu} (x^{\mu}) \rightarrow \overline{g}_{\mu \nu} (x^{\mu}) = \Omega (x^{\mu})^2 g_{\mu \nu} (x^{\mu}) \;.
\end{equation} 
\noindent Locally, the metric on the conformal diagram is then conformally equivalent to the metric of the space-time depicted. Under such a conformal rescaling, the causal nature of a curve or vector field remains invariant: a vector field is space-like with respect to $\overline{g}_{\mu \nu} (x^{\mu})$ if it is space-like with respect to $g_{\mu \nu} (x^{\mu})$; a curve is time-like with respect to $\overline{g}_{\mu \nu} (x^{\mu})$ if it is time-like with respect to $g_{\mu \nu} (x^{\mu})$. 

\par With these underlying ideas in place, let us proceed to the construction of the diagram itself. We begin with the introduction of new coordinates, motivated by the motion of physical particles. These are the advanced and retarded Eddington-Finkelstein coordinates, defined respectively as
\begin{equation}
u = t - r_* \quad \quad \text{and} \quad \quad v = t + r_* \;.
\end{equation} 
\noindent Here, $t$ and $r_*$ represent the Schwarzschild time coordinate and the tortoise coordinate, respectively. If we treat $v$ as a new time coordinate, we can rewrite the metric in the form of ingoing Eddington-Finkelstein coordinates,
\begin{equation} \label{eq:inEFcoords}
ds^2 = -\left(1- \frac{2M}{r} \right) dv^2 + 2dvdr + r^2 d\Omega^2_{\mathbb{S}^2} \;.
\end{equation}

\par Upon rewriting the $(t,r)$-component of the Schwarzschild metric in terms of both $v$ and $u$, we obtain
\begin{equation} \label{eq:dudv}
ds^2 = -\left(1- \frac{2M}{r} \right) du dv \;,
\end{equation}
\noindent with $r=r(u,v)$. Radial light rays correspond to $du \; dv =0$ with ingoing light rays following curves of constant $v$, while outgoing light rays follow curves of constant $u$. For $r > 2M$, outgoing light rays move towards larger $r$, while for $r <2M$, they move towards smaller $r$. For this reason, any observer following a time-like worldline eventually hits the space-like singularity at $r =0$. With these asymptotic behaviours in mind, we can summarise the asymptotic regions of the space-time using the notation provided in Table \ref{table:conformalconvention}. 

\par Let us now proceed to the construction of the conformal diagram. To do so, we introduce the Kruskal coordinates:
\begin{eqnarray}
X = \frac{1}{2} \left( e^{v/4M} + e^{-u/4M} \right) & = & (r-2M)^{1/2} e^{r/4M} \cosh \frac{t}{4M} \;, \\
T = \frac{1}{2} \left( e^{v/4M} - e^{-u/4M} \right) & = & (r-2M)^{1/2} e^{r/4M} \sinh \frac{t}{4M} \;,\\
\hspace{3cm} X^2 - T^2 & = & (r-2M)^{1/2} e^{r/2M} \;.
\end{eqnarray}
Here, $e^{v/4M}$ and $e^{-u/4M}$ are affine parameters along the outgoing  and ingoing null geodesics, respectively. We may then rewrite Eq. (\ref{eq:dudv}) as
\begin{equation}  \label{eq:SchwarzKruskal}
ds^2 = \frac{32M^3}{r} e^{-r/2M} (-dT^2 + dX^2) \;.
\end{equation}
\noindent Due to the symmetries of the $(X,T)$-plane, each surface of constant $r$ is represented twice, that is, two hyperbolas have the same value of $r$. As a result, there are event horizons at $T=\pm X$, an asymptotic region for $X \gg 0$ and for $X \ll 0$, as well as an $r=0$ singularity for $T>0$ and for $T<0$. These symmetries prove useful in the maximally-extended conformal diagram presented in Fig. \ref{fig:SchwarzConform}. 

\par With a conformal rescaling of Eq. (\ref{eq:SchwarzKruskal}), we can generate Fig. \ref{fig:SchwarzConform}. Regions I and II are covered by the ingoing Eddington-Finkelstein coordinates of Eq. (\ref{eq:inEFcoords}), where Region II is the black hole interior and Region I is the exterior asymptotically flat space-time. Here, $\mathscr{H}^+_R$ and $\mathscr{H}^-_R$ represent the future and past event horizons, respectively; they intersect at the bifurcation two-sphere $\mathscr{S}$. Hence, $\mathscr{H}^+_R$ and $\mathscr{H}^-_R$ are each branches of a bifurcate Killing horizon, and share an equivalent surface gravity $\kappa_+$.

\par When we discuss black holes formed from gravitational collapse, we consider only Regions I and II. Like Region I, Region IV is asymptotically flat, but it is causally disconnected from Region I. Region III is referred to as a \enquote{white hole}, so called as it is an antithesis to a black hole and is a hypothetical object from which information can escape but cannot enter. Regions III and IV are therefore treated as unphysical. 

\begin{table}[t]
\caption{\textit{Notation for asymptotic regions depicted in a standard conformal space-time diagram.  \label{table:conformalconvention}}}
\begin{center}
\begin{tabular}{@{}CCCC@{}}
\hline\noalign{\smallskip}
\multirow{2}{*}{Region} & \multirow{2}{*}{Symbol} & \multicolumn{2}{C}{Coordinates}\\
                   &                    & Schwarzschild & Eddington{\text{-}}Finkelstein \quad \\
\noalign{\smallskip}\hline\noalign{\smallskip}
 \text{Past horizon} & \mathscr{H}^- & t \rightarrow -\infty, \;r_* \rightarrow -\infty & v \rightarrow -\infty, \;\text{finite }u   \\
 \text{Future horizon} & \mathscr{H}^+ & t \rightarrow +\infty, \;r_* \rightarrow -\infty & \text{finite }v, \;u\rightarrow +\infty\\
 \text{Past null infinity} & \mathscr{I}^- & t \rightarrow -\infty, \; r_* \rightarrow +\infty & \text{finite }v, \; u\rightarrow-\infty\\
 \text{Future null infinity} & \mathscr{I}^+ & t \rightarrow +\infty, \; r_* \rightarrow +\infty & v \rightarrow +\infty, \; \text{finite }u \\
 \hline\noalign{\smallskip}
\\
\end{tabular}
\end{center}
\end{table}

\bibliographystyle{iopart-num} 
\bibliography{zotero}

\providecommand{\newblock}{}
\begin{thebibliography}{10}
\expandafter\ifx\csname url\endcsname\relax
  \def\url#1{{\tt #1}}\fi
\expandafter\ifx\csname urlprefix\endcsname\relax\def\urlprefix{URL }\fi
\providecommand{\eprint}[2][]{\url{#2}}

\bibitem{OppenheimerSnyder1939_StarCollapse}
Oppenheimer J~R and Snyder H 1939 {\em Phys. Rev.\/} {\bf 56} 455--459

\bibitem{Landsman2022_PenroseContextualised}
Landsman K 2022 {\em Gen. Rel. Grav.\/} {\bf 54} 115 (\textit{Preprint}
  \eprint{2205.01680})

\bibitem{Penrose1964_CC1}
Penrose R 1965 {\em Phys. Rev. Lett.\/} {\bf 14} 57--59

\bibitem{Penrose1969_CC2-WCCdef}
Penrose R 1969 {\em Riv. Nuovo Cim.\/} {\bf 1} 252--276

\bibitem{Penrose1999_CC}
Penrose R 1999 {\em Journal of Astrophysics and Astronomy\/} {\bf 20} 233--248

\bibitem{Penrose1973_Blueshift}
Simpson M and Penrose R 1973 {\em Int. J. Theor. Phys.\/} {\bf 7} 183--197

\bibitem{HorowitzCrisford2017_WCCtesting}
Crisford T, Horowitz G~T and Santos J~E 2018 {\em Phys. Rev. D\/} {\bf 97}
  066005 (\textit{Preprint} \eprint{1709.07880})

\bibitem{Rossetti2023_SCCmath}
Rossetti F 2023 {\em Ann. Henri Poincaré\/} (\textit{Preprint}
  \eprint{2309.14420})

\bibitem{Sbierski:2015nta}
Sbierski J 2018 {\em J. Diff. Geom.\/} {\bf 108} 319--378 (\textit{Preprint}
  \eprint{1507.00601})

\bibitem{Chrysostomou:2024inc}
Chrysostomou A, Cornell A, Deandrea A, Noshad H and Park S~C 2025 {\em PoS\/}
  {\bf ICHEP2024} 782

\bibitem{Israel1968_NoHair-RN}
Israel W 1968 {\em Commun. Math. Phys.\/} {\bf 8} 245--260

\bibitem{Kramer2003_ExactEFEsols}
Stephani H, Kramer D, MacCallum M, Hoenselaers C and Herlt E 2003 {\em Exact
  Solutions of Einstein's Field Equations\/} 2nd ed Cambridge Monographs on
  Mathematical Physics (Cambridge: Cambridge University Press)

\bibitem{AntoniadisBenakli2020_WGCdS}
Antoniadis I and Benakli K 2020 {\em Fortschritte der Physik\/} {\bf 68}
  2000054 (\textit{Preprint} \eprint{2006.12512})

\bibitem{Hawking1974_HawkT}
Hawking S~W 1974 {\em Nature\/} {\bf 248} 30--31

\bibitem{MossMyers1998_CC}
Brady P~R, Moss I~G and Myers R~C 1998 {\em Phys. Rev. Lett.\/} {\bf 80}
  3432--3435 (\textit{Preprint} \eprint{gr-qc/9801032})

\bibitem{GibbonsHawking1977_BHTherm}
Gibbons G~W and Hawking S~W 1977 {\em Phys. Rev. D\/} {\bf 15} 2738--2751

\bibitem{Romans1991_ColdLukewarmRNdS}
Romans L~J 1992 {\em Nucl. Phys. B\/} {\bf 383} 395--415 (\textit{Preprint}
  \eprint{hep-th/9203018})

\bibitem{DiasReallSantos2018_SCC}
Dias O~J~C, Reall H~S and Santos J~E 2019 {\em Class. Quant. Grav.\/} {\bf 36}
  045005 (\textit{Preprint} \eprint{1808.04832})

\bibitem{Chrysostomou2023_RNdS}
Chrysostomou A, Cornell A~S, Deandrea A, Noshad H and Park S~C 2023
  (\textit{Preprint} \eprint{2310.07311})

\bibitem{Bousso1996_Nariai}
Bousso R 1997 {\em Phys. Rev. D\/} {\bf 55} 3614--3621 (\textit{Preprint}
  \eprint{gr-qc/9608053})

\bibitem{vanRiet2019_FLevapBHdS}
Montero M, Van~Riet T and Venken G 2020 {\em JHEP\/} {\bf 01} 039
  (\textit{Preprint} \eprint{1910.01648})

\bibitem{Brill1993_RNdSextrema}
Brill D~R and Hayward S~A 1994 {\em Class. Quant. Grav.\/} {\bf 11} 359--370
  (\textit{Preprint} \eprint{gr-qc/9304007})

\bibitem{Mann1995_ChargedBHpairs}
Mann R~B and Ross S~F 1995 {\em Phys. Rev. D\/} {\bf 52} 2254--2265
  (\textit{Preprint} \eprint{gr-qc/9504015})

\bibitem{Bousso1999_QuantumStructredS}
Bousso R 1999 {\em Phys. Rev. D\/} {\bf 60} 063503 (\textit{Preprint}
  \eprint{hep-th/9902183})

\bibitem{HawkingEllis1973_LSSUbook}
Hawking S~W and Ellis G~F~R 1973 {\em The Large Scale Structure of
  Space-Time\/} Cambridge Monographs on Mathematical Physics (Cambridge:
  Cambridge University Press)

\bibitem{BritoCardosoPani2020_Superradiance}
Brito R, Cardoso V and Pani P 2020 {\em Superradiance\/} (Springer
  International Publishing) (\textit{Preprint} \eprint{1501.06570})

\bibitem{refBertiCardoso}
Berti E, Cardoso V and Starinets A~O 2009 {\em Class. Quant. Grav.\/} {\bf 26}
  163001 (\textit{Preprint} \eprint{0905.2975})

\bibitem{KonoplyaZhidenko2014_RNdSrprc}
Konoplya R~A and Zhidenko A 2014 {\em Phys. Rev. D\/} {\bf 90} 064048
  (\textit{Preprint} \eprint{1406.0019})

\bibitem{Papantonopoulos2022}
Gonz{\'a}lez P~A, Papantonopoulos E, Saavedra J and V{\'a}squez Y 2022 {\em
  JHEP\/} {\bf 06} 150 (\textit{Preprint} \eprint{2204.01570})

\bibitem{Wald1984_GRbook}
Wald R~M 1984 {\em General Relativity\/} (Chicago, USA: Chicago Univ. Pr.)

\bibitem{Hawking1966_Singularities}
Hawking S 2014 {\em Eur. Phys. J. H\/} {\bf 39} 403--411

\bibitem{Penrose1973_CC3}
Penrose R 1973 {\em Annals N. Y. Acad. Sci.\/} {\bf 224} 125--134

\bibitem{Dafermos:2012np}
Dafermos M 2014 {\em Commun. Math. Phys.\/} {\bf 332} 729--757
  (\textit{Preprint} \eprint{1201.1797})

\bibitem{Dafermos2018_BlueShiftLambda}
Dafermos M and {Shlapentokh-Rothman} Y 2018 {\em Class. Quant. Grav.\/} {\bf
  35} 195010 (\textit{Preprint} \eprint{1805.08764})

\bibitem{Christodoulou2008_SCC}
Christodoulou D 2008 The formation of black holes in general relativity {\em
  12th Marcel Grossmann Meeting on General Relativity\/} pp 24--34
  (\textit{Preprint} \eprint{0805.3880})

\bibitem{Gravitation1973}
Misner C~W, Thorne K~S and Wheeler J~A 1973 {\em Gravitation\/} (San Francisco:
  W. H. Freeman) ISBN 978-0-7167-0344-0

\bibitem{Dafermos:2017dbw}
Dafermos M and Luk J 2017  (\textit{Preprint} \eprint{1710.01722})

\bibitem{Dafermos:2014jwa}
Dafermos M, Rodnianski I and {Shlapentokh-Rothman} Y 2014  (\textit{Preprint}
  \eprint{1412.8379})

\bibitem{Teukolsky1973_KerrStability-I}
Teukolsky S~A 1973 {\em Astrophys. J.\/} {\bf 185} 635--647

\bibitem{TeukolskyPress1973_KerrStability-II}
Press W~H and Teukolsky S~A 1973 {\em Astrophys. J.\/} {\bf 185} 649--674

\bibitem{Luk:2015qja}
Luk J and Oh S~J 2017 {\em Duke Math. J.\/} {\bf 166} 437--493
  (\textit{Preprint} \eprint{1501.04598})

\bibitem{Hintz2015_SCC}
Hintz P and Vasy A 2017 {\em J. Math. Phys.\/} {\bf 58} 081509
  (\textit{Preprint} \eprint{1512.08004})

\bibitem{Costa2016_SCC}
Costa J~L and Franzen A~T 2017 {\em Annales Henri Poincare\/} {\bf 18}
  3371--3398 (\textit{Preprint} \eprint{1607.01018})

\bibitem{Franzen:2014sqa}
Franzen A~T 2016 {\em Commun. Math. Phys.\/} {\bf 343} 601--650
  (\textit{Preprint} \eprint{1407.7093})

\bibitem{Dafermos:2003yw}
Dafermos M and Rodnianski I 2005 {\em Invent. Math.\/} {\bf 162} 381--457
  (\textit{Preprint} \eprint{gr-qc/0309115})

\bibitem{refIKchap6}
Ishibashi A and Kodama H 2011 {\em Prog. Theor. Phys. Suppl.\/} {\bf 189}
  165--209 (\textit{Preprint} \eprint{1103.6148})

\bibitem{Dafermos:2007jd}
Dafermos M and Rodnianski I 2007  (\textit{Preprint} \eprint{0709.2766})

\bibitem{Cardoso2018_QNMsSCC}
Cardoso V, Costa J~L, Destounis K, Hintz P and Jansen A 2018 {\em Phys. Rev.
  D\/} {\bf 98} 104007 (\textit{Preprint} \eprint{1808.03631})

\bibitem{DiasEperonReallSantos2018_SCC}
Dias O~J~C, Eperon F~C, Reall H~S and Santos J~E 2018 {\em Phys. Rev. D\/} {\bf
  97} 104060 (\textit{Preprint} \eprint{1801.09694})

\bibitem{DiasReallSantos2018_SCC_Rough}
Dias O~J~C, Reall H~S and Santos J~E 2018 {\em JHEP\/} {\bf 10} 001
  (\textit{Preprint} \eprint{1808.02895})

\bibitem{Lagos2020_Anomalous}
Lagos M, Ferreira P~G and Tattersall O~J 2020 {\em Phys. Rev. D\/} {\bf 101}
  084018 (\textit{Preprint} \eprint{2002.01897})

\bibitem{Hod2023_ColdSCC}
Hod S 2023 {\em Int. J. Mod. Phys. D\/} {\bf 32} 2341004 (\textit{Preprint}
  \eprint{2305.08918})

\bibitem{Laplace1902_Probabilities}
Laplace P~S 1902 {\em A Philosophical Essay on Probabilities\/} (New York: J.
  Wiley \& Sons New York)

\bibitem{TimpeGriffithLevy2017_FreeWill}
Timpe K, Griffith M and Levy N (eds) 2017 {\em Routledge Companion to Free
  Will\/} (New York: Routledge)

\bibitem{Dafermos2003_RNinterior}
Dafermos M 2005 {\em Commun. Pure Appl. Math.\/} {\bf 58} 0445--0504
  (\textit{Preprint} \eprint{gr-qc/0307013})

\bibitem{Smeenk2021_DeterminismGR}
Smeenk C and Wuthrich C 2021 {\em Phil. Sci.\/} {\bf 88} 638--664
  (\textit{Preprint} \eprint{2009.07555})

\bibitem{Lawrence2022_PDEs}
Evans L~C 2022 {\em Partial Differential Equations\/} (Rhode Island: American
  Mathematical Society)

\bibitem{Nature_BlackHoleDefinitions}
Curiel E 2019 {\em Nature Astron.\/} {\bf 3} 27--34 (\textit{Preprint}
  \eprint{1808.01507})

\bibitem{Wald1995_BHthermQFTbook}
Wald R~M 1995 {\em Quantum Field Theory in Curved Space-Time and Black Hole
  Thermodynamics\/} Chicago Lectures in Physics (Chicago, IL: University of
  Chicago Press) ISBN 978-0-226-87027-4

\bibitem{Carter1973_Rigidity}
Carter B 1973 {Black holes equilibrium states} {\em {Les Houches Summer School
  of Theoretical Physics}: {Black Holes}\/} pp 57--214

\bibitem{refHorowitzChap1}
Horowitz G~T 2012 Black holes in four dimensions {\em Black Holes in Higher
  Dimensions\/} ed Horowitz G~T (Cambridge, UK: Cambridge University Press) pp
  3--20 1st ed

\end{thebibliography}
\end{document}